\documentclass[fleqn,usenatbib]{mnras}

\usepackage{amssymb}
\usepackage{newtxtext,newtxmath}
\usepackage{amsfonts}

\usepackage[T1]{fontenc}
\usepackage{ae,aecompl}
\usepackage{float}
\usepackage{comment}
\usepackage{xcolor}
\usepackage{upgreek}
\usepackage{subcaption}

\usepackage{grffile}
\usepackage{etoolbox}
\makeatletter 
  \patchcmd{\NAT@citex}
    {\@citea\NAT@hyper@{%
      \NAT@nmfmt{\NAT@nm}%
      \hyper@natlinkbreak{\NAT@aysep\NAT@spacechar}{\@citeb\@extra@b@citeb}%
      \NAT@date}}
    {\@citea\NAT@nmfmt{\NAT@nm}%
    \NAT@aysep\NAT@spacechar\NAT@hyper@{\NAT@date}}{}{}

  \patchcmd{\NAT@citex}
    {\@citea\NAT@hyper@{%
      \NAT@nmfmt{\NAT@nm}%
      \hyper@natlinkbreak{\NAT@spacechar\NAT@@open\if*#1*\else#1\NAT@spacechar\fi}%
        {\@citeb\@extra@b@citeb}%
      \NAT@date}}
    {\@citea\NAT@nmfmt{\NAT@nm}%
    \NAT@spacechar\NAT@@open\if*#1*\else#1\NAT@spacechar\fi\NAT@hyper@{\NAT@date}}
    {}{}
\makeatother

\usepackage{etoolbox}
 \makeatother
\usepackage{graphicx}	
\usepackage{amsmath}	
\usepackage{amssymb}	
\usepackage{textcomp}
\usepackage{pifont}
\title[Black Holes in the Early Universe]{Distinguishing the impact and signature of black holes from different origins in early cosmic history}

\author[ Zhang et al.]{
Saiyang Zhang,\textsuperscript{\href{https://orcid.org/0000-0003-1541-177X}{\includegraphics[width=2.5mm]{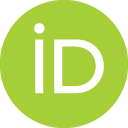}}\,}$^{1,2}$\thanks{E-mail: szhangphys@utexas.edu}
Boyuan Liu\textsuperscript{\href{https://orcid.org/0000-0002-4966-7450}{\includegraphics[width=2.5mm]{orcid.png}}\,}$^{3}$ and
Volker Bromm\textsuperscript{\href{https://orcid.org/0000-0003-0212-2979}{\includegraphics[width=2.5mm]{orcid.png}}}$^{2,4}$
\\
$^{1}$Department of Physics, University of Texas, Austin, TX 78712, USA\\ 
$^{2}$Weinberg Institute for Theoretical Physics, University of Texas, Austin, TX 78712, USA\\
$^{3}$Institute of Astronomy, University of Cambridge, Madingley Road, Cambridge, CB3 0HA, UK\\
$^{4}$Department of Astronomy, University of Texas, Austin, TX 78712, USA
}

\date{Accepted XXX. Received YYY; in original form ZZZ}

\pubyear{2023}

\begin{document}


\maketitle
\label{firstpage}
\pagerange{\pageref{firstpage}--\pageref{lastpage}}

\begin{abstract}
We semi-analytically model the effects of primordial black hole (PBH) accretion on the cosmic radiation background during the epoch of reionization ($z\gtrsim 6$). PBHs in the intergalactic medium (IGM) and haloes, where star formation can occur, are considered. For stars with a mass $\gtrsim 25 \rm\ M_{\odot}$, formed in suitable host haloes, we assume they quickly burn out and form stellar remnant black holes (SRBHs). These SRBHs, like PBHs, also accrete material, and are considered to have similar radiation feedback in the halo environment. To improve the background radiation estimation, we consider the impact of PBHs on structure formation, allowing a modified halo mass function. We consider the radiation feedback from a broad suite of black holes: PBHs, SRBHs, high-mass X-ray binaries (HMXBs), and supermassive black holes (SMBHs). The result shows that at $z\gtrsim 30$, the radiation background energy density is generated by PBHs accreting in the IGM, whereas at lower redshifts, halo accretion dominates. We also dissect the total power density by modeling the accretion spectral energy distribution (SED) across different wavebands. In the UV band, we find that for $f_{\rm PBH} \lesssim 10^{-3}$, the H-ionizing and Lyman-$\alpha$ fluxes from PBH accretion feedback do not violate existing constraints on the timing of reionization, and on the effective Wouthuysen-Field coupling of the 21-cm spin temperature of neutral hydrogen to the IGM kinetic temperature. However, in the X-ray band, with the same abundance, PBHs contribute significantly and could explain the unresolved part of the cosmic X-ray background.
\end{abstract}




\section{Introduction}

Dark matter (DM) has been a long-standing mystery in astrophysical cosmology \citep[e.g.][]{Feng2010}. Although it comprises a significant portion of the cosmic mass-energy density, we still have no knowledge of its micro-physical nature and origin. Whereas most explanations invoke extensions of the standard model of particle physics \citep[e.g.][]{Bertone2012}, an alternative DM scenario is based on primordial black holes (PBHs), formed via the collapse of overdense regions in the early Universe, prior to matter-radiation equality (\citealt{Zeldovich1967SvA....10..602Z, hawking1971gravitationally,Sureda2021MNRAS.507.4804S}). The recent detection of gravitational wave (GW) events (\citealt{Abbott2016PhysRevLett.116.061102, Abbott2020PhRvL.125j1102A}) from merging binary black holes (BBHs) has triggered a new wave of interest in PBHs. In particular, the detection of BBH merger events with total mass $\sim$150\,$\mathrm{M}_{\odot}$ (\citealt{Abbott2020PhRvL.125j1102A}) challenges current stellar evolution models \citep[e.g.][]{Liu_GW_2020}, raising the possibility of a primordial origin \citep[e.g.][]{Bird2016PhRvL.116t1301B,Kashlinsky2016ApJ...823L..25K,Clesse2017PDU....15..142C,Deluca2020JCAP...06..044D,Deluca2020JCAP...11..028D}.

However, it is unlikely that PBHs within the mass range $M_{\mathrm{PBH}}\sim 1-100\ \mathrm{M}_{\odot}$, typical for stars and their remnants, provide all of the cold dark matter (CDM) measured in the Universe, based on multiple empirical constraints \citep[e.g.][]{Carr2020ARNPS..70..355C}. Their contribution can be expressed as $f_{\mathrm{PBH}}=\Omega_{\mathrm{PBH}} / \Omega_{\mathrm{DM}}$, the ratio of mass fraction in PBH DM to the total CDM. A prominent observational signature is the GW signal generated from mergers of BBHs. The statistical study of BH merger rates captured by the 
LIGO-Virgo-KAGRA network (LVK; \citealt{Ali-Haimoud2017PhysRevDMerger, Abbott2021ApJ...913L...7A}) constrains the abundance of PBHs to within $f_{\mathrm{PBH}}\lesssim 0.1-10^{-3}$. Furthermore, the non-detection of the stochastic gravitational wave background (SGWB) by LVK gives a similar but weaker constraint of $f_{\mathrm{PBH}}\lesssim 0.1$ (\citealt{VaskonenPhysRevD.101.043015,Bagui2022PDU....3801115B}). The recent evidence for a SGWB from pulsar timing array data \citep[e.g.][]{NGRAV2023} implicates a cosmological population of supermassive black hole (SMBH) binaries, and provides promising constraints on supermassive PBHs ($f_{\rm PBH}\lesssim 0.01$) as well \citep[e.g.][]{Acuna2023,Gouttenoire2023}. 

Even though stellar-mass and supermassive PBHs only make up a small fraction of dark matter, they can play important roles in early cosmic history. 
One of the main physical processes that could affect cosmic evolution is PBH accretion, as the focus of this paper. 
When PBHs travel through gaseous environments, by the pull of their gravitational well, they will attract and accrete the surrounding gas. In this study, for most cases, we will assume steady accretion of gas onto BHs, as found in \cite{Bondi1952MNRAS.112..195B}. The accreted gas will lose potential energy and emit the heat generated during this process. The resulting feedback radiation will affect the surrounding medium that PBHs are located in (e.g. \citealt{Takhistov2022JCAP}), and will change the overall cosmic thermal history (e.g. \citealt{Mena2019PhysRevD.100.043540}). More specifically, multiple studies show that the energy feedback from this accretion will contribute to cosmic radiation backgrounds at different wavebands (e.g. \citealt{Ali-Haimoud2017PhRvD,Inoue2017JCAP,DeLuca2020prd,DeLuca2022PhLB,Cappelluti2022ApJ,Ziparo2022MNRAS}). The change in cosmic microwave background (CMB) power spectra derived from PBH accretion limits their abundance, where stellar-mass PBHs are found to comprise no more than a small fraction of DM, $f_{\mathrm{PBH}}\lesssim 0.1-10^{-3}$ (e.g. \citealt{Poulin2017PhRvD, Serpico2020PhRvR}), similar to the constraints derived from GW observations. Further constraints from accretion onto PBHs are summarized in \cite{Carr2021}.

Moreover, with the recently launched JWST, observing in the near-infrared (NIR) electromagnetic (EM) wave band [$0.7-5\ \mathrm{ \mu m}$], astronomers start to unveil an increasing number of galaxies at $z \gtrsim 12$ (e.g. \citealt{Finkelstein2022ApJ...940L..55F}), discover SMBHs during their early formation stages 
\citep[e.g.][]{Natarajan2017ApJ...838..117N,Larson2023ApJ...953L..29L}, and aim to study anisotropies in the cosmic near-infrared background (CIB; \citealt{Kashlinsky2015ApJ...804...99K,Helgason2016MNRAS.455..282H}). Some of the galaxies, if confirmed by spectroscopic follow-up, are so massive that they could challenge the standard $\Lambda$CDM model of cosmological structure formation (\citealt{Labbe2023Natur.616..266L,Boylan2023}). PBHs can potentially address these latest observations in a broader context with their effects on the matter power spectrum and the subsequent structure formation history, as well as on the halo mass distribution  (e.g. \citealt{Afshordi2003ApJ...594L..71A,Carr2018MNRAS.478.3756C,Kashlinsky2021PhRvL.126a1101K,AtrioBarandela2022ApJ...939...69A_PBHStreaming,Boyuan2022ApJ}). This change to the halo mass function will also affect the total accretion feedback from PBHs located in haloes, and consequently the cosmic radiation background generated by the aggregate emission (\citealt{Kashlinsky2016ApJ...823L..25K,Cappelluti2022ApJ,Boyuan2022MNRAS.514.2376L}).

In this paper, we will assess the impact of PBHs in the mass range $M_{\rm PBH}\sim 1-100\ \rm M_{\odot}$ on early cosmic history, considering all BHs generated from different mechanisms as possible power sources contributing to the cosmological background radiation. We will consider the redshift range $z\sim 100 -6$, starting from an almost homogeneous matter distribution to the end of reionization. Overall, we will divide our treatment into PBHs floating in the intergalactic medium (IGM), and those residing in DM haloes. Calculating the resulting feedback energy from the accretion of the surrounding gas, in the former case we consider the effects of Hubble flow and Compton scattering, which will effectively slow down the accretion process \citep[see][]{Ricotti2007ApJI,Ricotti2008ApJII}. For the latter case, for simplicity, we assume that PBHs sit near the center of the halo following an isothermal density profile, as previously simulated in \cite{Boyuan2022MNRAS.514.2376L}. 
In addition, we consider astrophysical BHs generated from the death of massive stars ($\gtrsim 30 \mathrm{\,M}_{\odot}$) in virialized haloes \citep[e.g.][]{Husain2021MNRAS}, as well as SMBHs in more massive haloes with $M_{\mathrm{h}} \gtrsim 10^8 \mathrm{\,M}_{\odot}$ (e.g. \citealt{Jeon2022MNRAS}). This broad suite of accreting BHs in the early Universe is likely contributing to the sources for the unresolved part of the fluctuations in the cosmic X-ray and infrared backgrounds (CXB/CIB), as summarized in the review by \citet{Kashlinsky2018RvMP...90b5006K}.

A similar topic has been discussed in recent works \citep{Cappelluti2022ApJ,Ziparo2022MNRAS}, where the radiation background from PBH accretion has been evaluated. However, different from the assumptions of those studies, we will focus on the radiation background contribution from all BH sources, including stellar-remnant black holes (SRBHs), such as high-mass X-ray binaries (HMXBs). Therefore, we will integrate over haloes with a broad mass range of $\sim 10^5-10^{13} \mathrm{\,M}_{\odot}$, whereas \cite{Ziparo2022MNRAS} only considered minihaloes without star formation ($\lesssim 10^7 \mathrm{\,M}_{\odot}$).   Also, we will consider the case of enhanced accretion by PBH-seeded DM haloes within the IGM, whereas \cite{Cappelluti2022ApJ} only considered PBH accretion in star-forming haloes.
With our new implementations, we will reach a broader understanding of the PBH-feedback-generated radiation backgrounds. 

The structure of our paper is as follows. Section~\ref{sec:BHs} will address the impact of PBHs on structure formation and introduce the different sources of black holes, to be considered in the following treatment. In Section~\ref{sec:accret}, we will discuss the accretion physics in the IGM and DM haloes separately. Subsequently, we derive the resulting generation of radiation and heat, assessing the corresponding observational signatures in Section~\ref{sec:radfed}. We summarize and offer conclusions in Section~\ref{sec:summary}. Throughout this paper, we adopt a PBH-DM $\Lambda$CDM cosmology with parameters: $\Omega_{\mathrm{m}}=0.3153$, $\Omega_{\mathrm{b}}=0.04930$,  $n_{\mathrm{s}}=0.96$, and $h=0.6736$ \citep{Plank2020A&A...641A...6P}. The density fraction of CDM is then given by: $\Omega_{\mathrm{DM}} = \Omega_{\mathrm{m}} - \Omega_{\mathrm{b}}$, with a possible PBH sub-component of $\Omega_{\rm PBH}=f_{\rm PBH}\Omega_{\rm DM}$.

\vspace{-3ex}
\section{Black Holes in the Early Universe} \label{sec:BHs}
\begin{figure*}
    \centering
    \includegraphics[width=1.9\columnwidth]{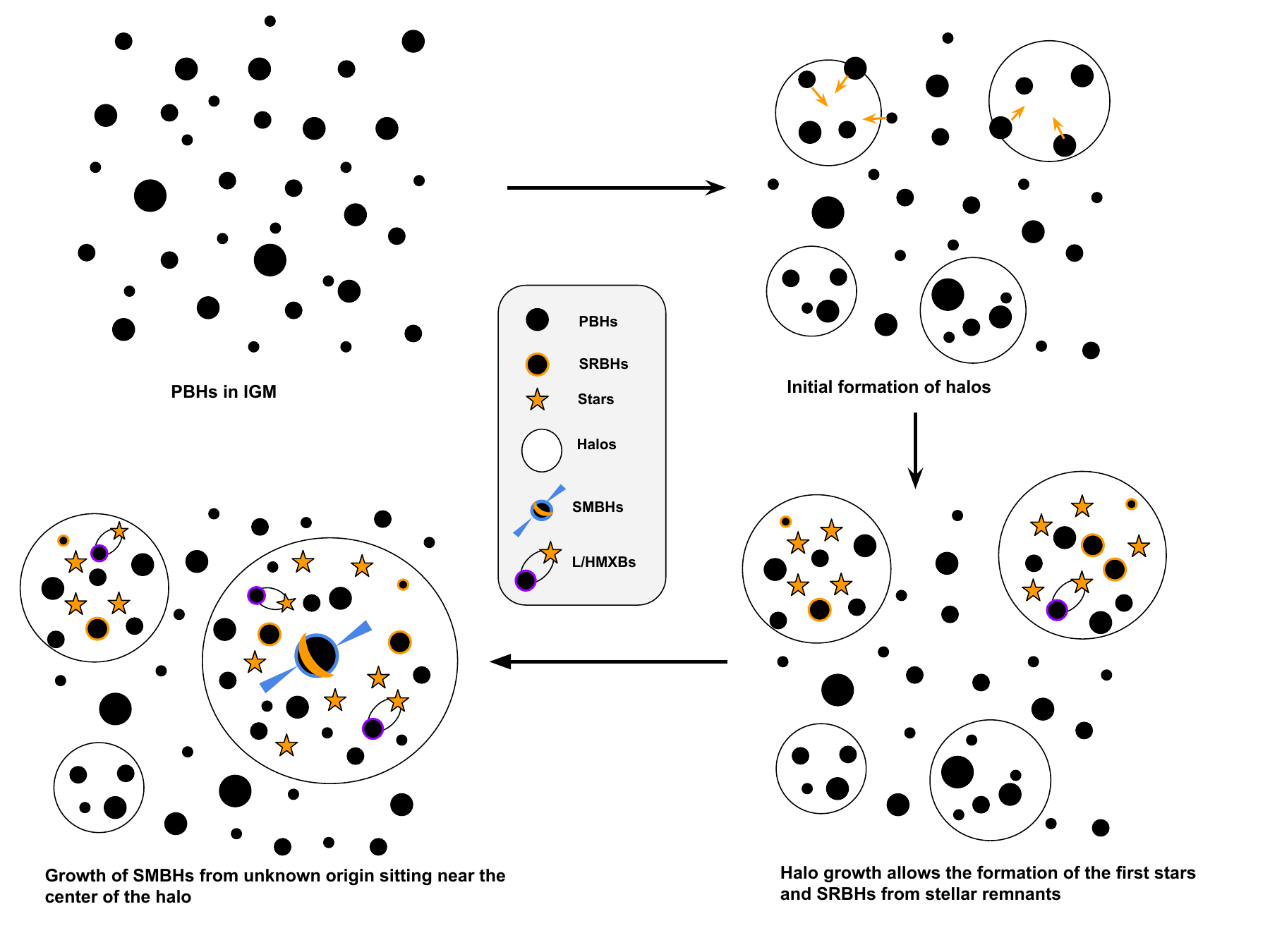}
    \vspace{-20pt}
    \caption{Overall role of PBHs in early cosmic evolution. Initially, before the onset of structure formation, PBHs reside in the quasi-homogeneous intergalactic medium (IGM), as shown in the \textbf{top left panel}. As the Universe expands and cools, haloes of small masses ($\sim 10^5-10^6 \mathrm{\, M}_{\odot}$) start to emerge from the gravitational instability, and PBHs cluster in those minihaloes, as illustrated in the \textbf{top right panel}. As haloes keep growing, the first stars form in sufficiently massive haloes from the collapse of primordial gas clouds. Stars with mass below the pair-instability limit ($\sim 10-130 \mathrm{\, M}_{\odot}$) quickly die and collapse into stellar-remnant black holes (SRBHs). In certain cases, some of the stars form in binaries, where the more massive component becomes a black hole. Through accretion from the companion star, the black hole emits strong X-rays, giving rise to low/high mass X-ray binaries (L/HMXBs), depending on the mass of the BH (see the \textbf{lower right panel}). As some of the haloes have grown increasingly massive, in excess of $\gtrsim 10^8 \mathrm{\, M}_{\odot}$, the black hole in the center of the galactic bulge starts to accrete at nearly Eddington rate to grow into supermassive black holes (SMBHs), which exert vigorous radiative feedback during this process. This more complex stage in cosmic evolution, with the full panoply of BH sources, is summarized in the \textbf{lower left panel}.   }
    \label{fig:evoscheme}
\end{figure*}

In this section, we will address the overall cosmic evolution at high redshifts in the presence of PBHs, as summarized in Figure~\ref{fig:evoscheme}, where we display the entire suite of BHs from different origins. First, in Section~\ref{sec:CS}, we briefly summarize previous results on how cosmological structure formation is modified by PBH DM, including changes to the  halo mass function, as compared to standard $\Lambda$CDM. Then, in Section~\ref{sec:PBHs}, we touch on 
select PBH formation mechanisms and the resulting mass distribution. We similarly discuss relevant aspects for BHs of astrophysical origins in Section~\ref{sec:ABHs}. 
\vspace{-10pt}
\subsection{Cosmological Structure Formation}
\label{sec:CS} 
  Following the general formalism introduced in previous work (\citealt{Afshordi2003ApJ...594L..71A,Inman2019PhRvD.100h3528I}), we consider density perturbations as a function of redshift $z$, consisting of adiabatic and isocurvature terms, $\delta_{\mathrm{ad}}(z)$ and  $\delta_{\text{iso }}(z)$, respectively. Adiabatic perturbations uniformly affect all mass-energy components, such that $\delta_{\mathrm{ad}}(z)$ applies to all forms of DM. In this context, the isocurvature term derives from the discreteness of individual PBHs (\citealt{Kashlinsky2016ApJ...823L..25K,Desjacques2018PhRvD..98l3533D,Deluca2020JCAP...11..028D}), and can be written as
\begin{equation}
\delta_{\mathrm{iso}}(\mathbf{x})\equiv \frac{\delta \rho_{\rm PBH}}{\Bar{\rho}_{\rm DM}}=\frac{f_{\rm PBH}}{\bar{n}_{\mathrm{PBH}}} \sum_i \delta_D\left(\mathbf{x}-\mathbf{x}_i\right)-1.
\end{equation}
Here, $\delta_D $ is the Dirac delta function, $\mathbf{x}_i$ marks the position of the $i$-th PBH, and $\bar{n}_{\mathrm{PBH}}$ is the average PBH number density, which depends on the average PBH mass. This formalism can be extended to a broader mass spectrum, with a characteristic mass scale of $M_{\rm c}$, as stated in, e.g., \citet{Carr1975ApJ...201....1C,Choptuik1993PhRvL..70....9C}. In general, we thus write the average density as
\begin{equation}
\bar{n}_{\mathrm{PBH}}  =\frac{f_{\mathrm{PBH}} }{M_{\rm c}} \Omega_{\mathrm{DM}} \frac{3 H_0^2}{8 \pi G}\mbox{\ .}
\end{equation}
  
Given the limits on the abundance of PBHs, implying that they do not compose the totality of DM, we assume that the remainder of (cold) DM is provided by particle DM (PDM). We thus consider an additional term imprinted by the PBH isocurvature perturbations on the PDM, such that the combined evolution of the density perturbations in both DM (PDM) and PBHs is given as follows:
\begin{equation}
\begin{aligned}
 \delta_{\mathrm{DM}}(z) & = \delta_{\mathrm{ad}}(z)+\delta_{\mathrm{iso}}(z) =\delta_{\mathrm{ad}}(z)+\left[T_{\mathrm{iso}}(z)-1\right] f_{\mathrm{PBH}} \delta_{\text {iso }}^0, \\
\delta_{\mathrm{PBH}}(z) & =\delta_{\mathrm{DM}}(z)+\delta_{\text {iso }}^0=\delta_{\mathrm{ad}}(z)+T_{\text {iso }}^{\mathrm{PBH}}(z) \delta_{\text {iso }}^0.\\
\end{aligned}
\end{equation}
The evolution of adiabatic perturbations with respect to $z$ is expressed via $\delta_{\mathrm{ad}}(z) = T_{\mathrm{ad}}(z)\delta_{\rm ad}^{0}$. Here, $\delta_{\rm ad}^{0}$ and $\delta_{\rm iso}^{0}$ correspond to the primordial adiabatic and PBH isocurvature perturbations, whereas
 $T_{\rm ad}(z)$ and $T_{\rm iso}(z)$ are the linear transfer functions for the two modes, both normalized to 1 as $z\rightarrow \infty$. The perturbation in PDM induced by PBHs is $\delta_{\rm iso}(z)= \left[T_{\rm iso}(z)-1\right]f_{\rm PBH}\delta_{\rm iso}^{0}$, and we can write the transfer function for PBH-induced isocurvature fluctuations as $T^{\rm PBH}_{\rm iso}(z)=1+\left[T_{\rm iso}(z)-1\right]f_{\rm PBH}$. We note that the transfer function is in turn related to the structure growth factor, $T(z) = D(z)$. To a good approximation, we can use the results from solving the cosmic fluid equations with embedded point masses \citep{Meszaros1974A&A....37..225M}, and adopt the following analytical fit for the growth factor \citep{Inman2019PhRvD.100h3528I}:   
\begin{equation}
\begin{aligned}
D(z) & \approx\left(1+\frac{3 \gamma}{2 \alpha_{-}} s\right)^{\alpha_{-}}, \quad s=\frac{1+z_{\mathrm{eq}}}{1+z} \\
\gamma & =\frac{\Omega_m-\Omega_b}{\Omega_m}, \quad \alpha_{-}=\frac{1}{4}(\sqrt{1+24 \gamma}-1),
\end{aligned}
\end{equation}
where $z_{\mathrm{eq}} \sim 3400$ represents the redshift at matter-radiation equality. Since we restrict our treatment to the matter-dominated era with $z\ll z_{\mathrm{eq}}$, we have $s\gg 1 \ (z \rightarrow 0)$. Therefore, $D_{\mathrm{iso}} \approx  3s /2 $ (assuming $\gamma \approx 1$ and $\alpha_{-}\approx 1$ to be constant). 

 Given the two-point correlation function for the density perturbations from PBH DM:
 \begin{equation}
 \left\langle\delta_{\mathrm{PBH}}(\mathbf{x}) \delta_{\mathrm{PBH}}(\mathbf{0})\right\rangle =\frac{f_{\rm PBH}^2}{\bar{n}_{\mathrm{PBH}}} \delta_D(\mathbf{x})+\xi_{\mathrm{PBH}}(x),
\end{equation}  

where $\xi_{\mathrm{PBH}}(x)$ represents the reduced correlation function as a function of separation distance $x = |\mathbf{x}|$ \citep[e.g.][]{ChisholmPhysRevD.73.083504,Desjacques2018PhRvD..98l3533D,dizgah_primordial_2019,DeLuca2022PhRvL.129s1302D}, we can calculate the power spectrum by taking the spatial Fourier transform of the correlation function of all perturbation modes. Specifically, the isocurvature term $P_{\rm iso}(k)$ from PBHs approximates to $f_{\rm PBH}^{2}/\bar{n}_{\rm PBH}$. We can then express the total power spectrum\footnote{In our calculation, we have used the package Colossus \citep{Diemer2018ApJCOLOSSUS} to generate the $\Lambda$CDM power spectrum, $P_{\rm\Lambda CDM}(k)$, as calibrated by \cite{Plank2020A&A...641A...6P}.} at $z=0$ as follows:
\begin{equation}
 \begin{aligned}
    P_{\mathrm{tot}}(k)&=P_{\rm\Lambda CDM}(k) + P_{\rm iso}(k), \\
    P_{\rm iso}(k) &= \left[f_{\rm PBH}D_{0}\right]^{2}/\bar{n}_{\rm PBH} + T^{2}_{\rm mix}(k)P_{\rm \Lambda CDM}(k).
 \end{aligned}
\label{eq:pk}
\end{equation}
Initially, the adiabatic and iso-curvature modes are not correlated. However, at a later stage ($z\sim 10$), we need to consider the correlation between the isocurvature and adiabatic perturbations since PBHs are more abundant in overdense regions on larger scales, but disrupt DM structure around them on smaller ones. Here, similar to the prescription in \cite{Boyuan2022MNRAS.514.2376L}, we introduce a mode mixing term governed by the transfer function $T_{\mathrm{mix}}$.  Considering the inter-PBH separation scale, $k_{\mathrm{PBH}}=\left(2 \pi^2 \bar{n}_{\mathrm{PBH}}\right)^{1 / 3}$, we use the heuristic formula calibrated to simulation results in \citet{Boyuan2022MNRAS.514.2376L}:
 \begin{equation}\label{eq:mixing}
T_{\mathrm{mix}}^2(k)= \begin{cases}f_{\mathrm{PBH}} D_0^2 D_{\mathrm{ad}, 0}^{-1}\left(k / k_{\mathrm{PBH}}\right)^3, & k \leq 3 k_{\mathrm{PBH}} \\ 0, & k>3 k_{\mathrm{PBH}}\end{cases},
\end{equation}
where $D_{0}=D(z=0)$ is the growth factor for the isocurvature mode, and $D_{\rm ad,0}=D_{\rm ad}(z=0)/D_{\rm ad}(z=z_{\rm eq})$ for the adiabatic mode\footnote{We denote the growth factor of the adiabatic mode as $D_{\rm ad}(z)$, and adopt the semi-analytical fit from Equ.(4.74)-(4.76) in \cite{mo2010galaxy}.}, both evaluated at redshift 0.

\begin{figure}
\centering
\begin{subfigure}{0.5\textwidth}
    \includegraphics[width=\textwidth]{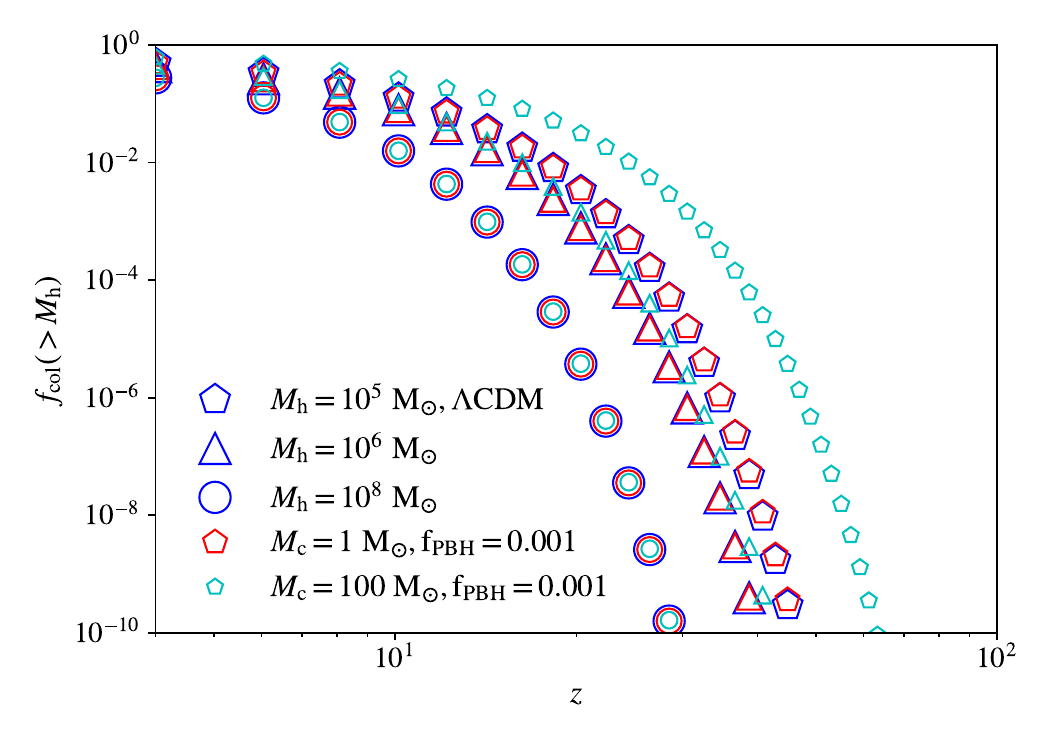}
    \vspace{-15pt}
    \caption{Constant mass fraction $f_{\rm PBH}=0.001$, but considering changes in mass scale: $M_{\rm c} = 1 \rm \ M_{\odot}$ (red) and $100 \rm \ M_{\odot}$ (cyan).}
    \label{fig:first}
\end{subfigure}
\begin{subfigure}{0.5\textwidth}
    \includegraphics[width=\textwidth]{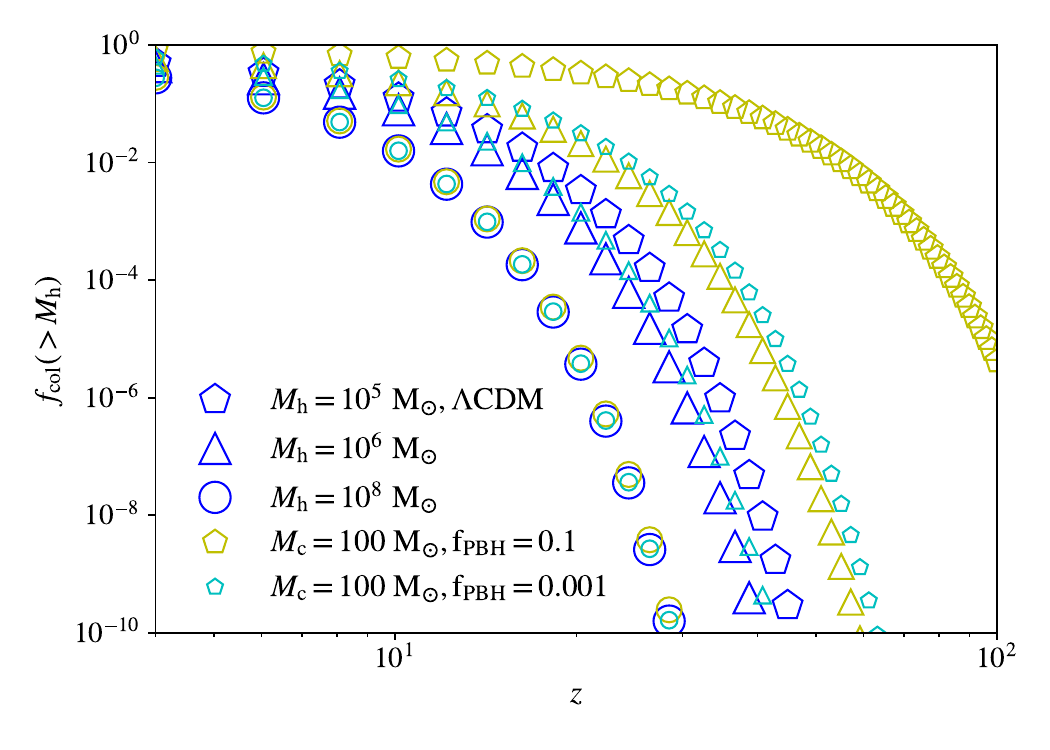}
    \vspace{-15pt}
    \caption{Varying PBH abundance, $f_{\rm PBH}=0.1 $ (yellow) and $0.001$ (magenta), keeping the mass scale constant at $M_{\rm c} = 100\rm \ M_{\odot}$.}
    \label{fig:second}
\end{subfigure}
\caption{Collapse fraction in haloes larger than $10^5$ (pentagon), $10^6$ (triangle), and $10^8 \rm M_{\odot}$ (circle), vs. redshift $z$. Here, we consider the changes in the formation of structures in the presence of PBHs (red, cyan and yellow, compared to the standard $\Lambda$CDM case (blue).}
\label{fig:fcol}
\end{figure}

 From the total power spectrum, we calculate the density contrast as a function of the window mass $M$:
\begin{equation}
    \sigma (M)=\left[\frac{1}{2 \pi^2} \int P_{\mathrm{tot}}(k) W_{\mathrm{TH}}\left(k r_{\rm M}\right) k^2 d k\right]^{1 / 2},
\end{equation}
where $W_{\mathrm{TH}}\left(k r_{\rm M}\right)$ is the top-hat Fourier transform of the spherical window function, and $r_{\rm M}$ corresponds to the comoving radius associated with the mass window $M$. We note that for non-linear scales, where $\sigma(M)\gtrsim 1$, this mass corresponds to a virialized halo structure. Therefore, in the following, we will assume $M_h=M$ for the halo mass.
Employing the Press-Schechter formalism for the halo distribution \citep{Press1974ApJ} \footnote{We note that the effect of PBH clustering at smaller scales, mainly represented by $\xi_{\mathrm{PBH}}(x)$, will contribute to the non-Gaussian part of the density field \citep[e.g.][]{ChisholmPhysRevD.73.083504,Desjacques2018PhRvD..98l3533D, Kashlinsky2016ApJ...823L..25K,AtrioBarandela2022ApJ...939...69A_PBHStreaming}. Our use of the standard Press-Schechter formalism here thus is a simplification, approximating the halo abundance. Another simulation study will be needed, in continuation of \citet{Boyuan2022MNRAS.514.2376L}, to fully address the effect of PBH clustering.}, the number density of virialized structures (halo mass function) between $M_{\rm h}$ and $M_{\rm h}+dM_{\rm h}$ is: 
\begin{equation} \label{eq:HMF}
    \begin{aligned}
n(M_{\rm h}, z) \mathrm{d} M_{\rm h} &=\sqrt{\frac{2}{\pi}} \frac{\bar{\rho}}{M_{\rm h}^2} \frac{\delta_{\mathrm{c}}(z)}{\sigma(M_{\rm h})} \\ &\times \exp \left(-\frac{\delta_{\mathrm{c}}^2(z)}{2 \sigma^2(M_{\rm h})}\right)\left|\frac{\mathrm{d} \ln \sigma(M_{\rm h})}{\mathrm{d} \ln M_{\rm h}}\right| \mathrm{d} M_{\rm h} ,
\end{aligned}
\end{equation}
where $\bar{\rho}$ is the average comoving matter density, and the critical overdensity factor, $\delta_{\mathrm{c}}(z)$, as a function of redshift is given by
\begin{equation}
    \delta_c(z) \simeq \frac{1.686}{D_{\rm ad}(z)}.
\end{equation} 
Integrating the halo number density over the mass range of interest and dividing by the average mass density, the collapse fraction of the cosmic matter is given by
\begin{equation}
    f_{\rm col}(>M_{\rm h}|z)=\frac{1}{2} \text{erfc} \left(\frac{\delta_c(z)}{\sqrt{2}\sigma(R)}\right),
\end{equation}
and is plotted in Figure~\ref{fig:fcol}.

As can be seen, in the presence of PBHs, more minihaloes ($M_{\rm h}\sim 10^5\rm \ M_{\odot}$) form than in the standard $\Lambda$CDM case. Also, the abundance of minihaloes will depend on the characteristic PBH mass, $M_{\rm c}$, but more strongly on the PBH abundance. This behaviour can be explained by the shift of total power spectrum $P_{\rm tot}$ in the presence of PBHs by the isocurvature term $f_{\rm PBH}^{2}/\bar{n}_{\rm PBH}$ in Equ.~(\ref{eq:pk}), which changes the density contrast subsequently. In addition, we have examined the effect of the mode mixing term in Equ.~(\ref{eq:mixing}), finding that it does not significantly change the collapse fraction within the PBH mass scale considered here. However, as we increase the mass scale to $\gtrsim 100 \rm \ M_{\odot}$, minihaloes of mass $\lesssim 10^6 \rm \ M_{\odot}$ would be less abundant, if mode mixing were excluded. Going forward, the collapse fraction allows us to calculate the PBH abundance in the IGM, whereas the halo mass function enables us to evaluate the prevalence of PBHs in DM haloes.

\vspace{-1ex}

\subsection{Primordial Black Holes}\label{sec:PBHs}
Several previous works provide reviews of the PBH formation mechanisms and abundance constraints from phenomenological studies \citep[e.g.][]{Carr2020ARNPS..70..355C,Carr2021,Escriva2023JCAP...05..004E}. In this paper, we consider PBHs with different mass distributions and study how these changes affect the radiation background output.
To begin, the PBH mass distribution $\psi(M_{\mathrm{PBH}})$ with respect to PBH mass $M_{\mathrm{PBH}}$ is related to the differential number density as follows:
\begin{equation}
    \frac{d n_{\mathrm{PBH}}}{d \ln M_{\mathrm{PBH}}}= f_{\mathrm{PBH}} \Omega_{\mathrm{DM}}\frac{3 H_0^2}{8 \pi G} \psi(M_{\mathrm{PBH}}).
\end{equation}
Here, $\psi(M_{\mathrm{PBH}})$ is the mass probability distribution function,  normalized to 1 when integrating over $\ln M_{\mathrm{PBH}}$, and it reflects the specific PBH formation mechanism.
In the most common scenario, PBHs follow a monochromatic mass distribution when all PBHs form from the collapse of overdensities during the radiation-dominated era \citep[e.g.][]{Carr1975ApJ...201....1C}. The mass function then takes the simple form:
\begin{equation}
     \psi\left(M_{\mathrm{PBH}}\right) = \delta (M_{\mathrm{PBH}}-M_{\rm c}),
\end{equation}
where $M_{\rm c}$ is the characteristic PBH mass scale. 

Another scenario suggests that PBHs form in smooth peaks of the inflation power spectrum, generating an initial mass function with a lognormal shape \citep{Dolgov:1992pu}. 
The functional form is given by:
\begin{equation}
    \psi\left(M_{\mathrm{PBH}}\right)=\frac{1}{\sqrt{2 \pi} \sigma M_{\mathrm{PBH}}} \exp \left(-\frac{\ln ^2\left(M_{\mathrm{PBH}} / M_{\rm c}\right)}{2 \sigma^2}\right),
\end{equation}

where $\sigma$ denotes the width of the PBH distribution. Constraints on the parameter space $(f_{\rm PBH},\sigma,M_{\mathrm{c}})$ for the lognormal distribution have been calculated in \cite{Kuhnel2017PhRvD}, specifically concluding that $ 0 < \sigma \leq 1$. We note that as $\sigma \rightarrow 0$, the lognormal distribution reduces to the monochromatic distribution. We also note that many other shapes of PBH mass functions are discussed in the literature, like the power-law or critical collapse functions \citep{Carr2021}. 
Here, we mostly restrict our attention to the monochromatic mass function, but our calculation also applies to lognormal distributions with a narrow dispersion. A comprehensive investigation with respect to different shapes of the initial PBH mass function will require another study.

\vspace{-10pt}
\subsection{Astrophysical Black Holes}\label{sec:ABHs}

\subsubsection{Supermassive Black Holes}
We will consider SMBHs as powerful sources of radiation in difference from other astrophysical black holes. For a comprehensive discussion of the formation and impact of SMBHs, we refer the reader to recent reviews \citep[e.g.][]{SMBH2019ConPh..60..111S,Inayoshi2020}. Our focus here is on the mass relationship between the SMBH and its host halo. For the sake of our analysis, we operate under the optimistic\footnote{Recent JWST observations have discovered a population of active galactic nuclei (AGN) at $z\gtrsim 5$ with a much higher abundance than previously expected, which implies a surprising ubiquity of SMBHs in high-$z$ galaxies \citep[e.g.][]{Furtak2023,Greene2023,Harikane2023,Kocevski2023,Maiolino2023,Matthee2023}. Motivated by this discovery, we assume that all haloes above a certain mass threshold host SMBHs.} assumption that SMBHs are located at the center of all massive haloes with $M_{\rm h} \gtrsim 10^8 \rm \ M_{\odot}$. This assumption allows us to directly derive the SMBH mass function from the halo mass function. Specifically, the mass of the central black hole is related to that of its host halo, as follows:
\begin{equation}
    M_{\rm SMBH}=\eta(M_{\rm h},z) M_{\rm h},
\end{equation}
where the BH formation efficiency, $\eta(M_{\rm h},z)$,  is a function of halo mass and redshift. In \cite{Jeon2022MNRAS}, it is calculated using a global energy balance argument as:
\begin{equation}
    \eta \simeq 2 \times 10^{-7}\left(\frac{1+z}{10}\right)\left(\frac{f}{0.1}\right)^{-1}\left(\frac{M_{\rm h}}{10^{10} \mathrm{M}_{\odot}}\right)^{2 / 3},
\end{equation}
where $f$ is the fraction of the BH rest mass energy deposited into the host halo gas as heat.
However, this idealized formalism scales as $\eta \propto (1+z)$,  and it does not converge to $0$ at high redshifts ($z \gtrsim 15$), where SMBHs begin to form. Therefore, we also adopt the fitted value of $\eta(M_{\rm h},z)$ provided by the UNIVERSE MACHINE, which is based on observational data \citep{Zhang2023MNRAS.518.2123Z}. This is discussed in more detail in Appendix \ref{AppendA}.

\vspace{-10pt}

\subsubsection{Isolated Black Holes from Stellar Remnants}

We also consider stellar-remnant black holes (SRBHs) as the eventual product of the steller evolution. The resulting abundance of SRBHs hinges on modeling of both the underlying stellar evolution and star-formation history. To proceed, we need to apply constraints on BHs derived from their progenitor stars. The eventual fate of a single star with an initial mass $M_{\mathrm{i}}$ can be summarized as follows \citep{Heger2003ApJ...591..288H}:
\begin{itemize}
    \item $M_{\mathrm{i}}<10 \mathrm{\,M}_{\odot} \rightarrow$ white dwarf 
    \item $10 \mathrm{\,M}_{\odot}<M_{\mathrm{i}}<M_{\star,\mathrm{c}} \rightarrow$ neutron star 
    \item $M_{\star,\mathrm{c}}<M_{\mathrm{i}}<140 \mathrm{\,M}_{\odot} \rightarrow$ black hole
    \item $140 \mathrm{\,M}_{\odot}<M_{\mathrm{i}}<260 \mathrm{\,M}_{\odot} \rightarrow$ pair-instability supernova (PISN) $\rightarrow$ no remnant
    \item $M_{\mathrm{i}}>260 \mathrm{\,M}_{\odot} \rightarrow$ massive black hole
\end{itemize}

Here, $M_{\star,\mathrm{c}}$ is the critical mass, before mass loss, above which a single star will eventually produce a BH, with a value of approximately  $25 \rm \ M_{\odot}$. According to \cite{Heger2003ApJ...591..288H}, the products of stellar evolution also depend on the metallicity of the stars. For extremely low metallicity stars, such as Population~III (Pop~III) stars, the summary above accurately captures the fate of dying stars. However, for higher metallicity Pop~I/II stars with $Z\gtrsim 10^{-4}\ \rm Z_{\odot}$, where mass loss is expected to be severe, the PISN stage of evolution may never occur, and is replaced by a BH remnant fate. Therefore, we assume that all Pop~I/II stars with an initial mass larger than the critical mass will evolve into BHs.

The literature offers various forms for the initial mass function (IMF) for different stellar populations, often characterized by a single or broken power-law \citep[e.g.][]{Chabrier2003PASP..115..763C}. Since our primary focus here is on the more massive stars that have the potential to form BHs, we adopt a simple power law for the stellar IMF:
\begin{align}\label{eq:IMF}
   \frac{d N}{d M_{\mathrm{i}}} =C M_{\mathrm{i}}^\alpha,
\end{align}
where $C$ is a normalization constant. For simplicity, we assume that any BH with a progenitor mass larger than $260 \mathrm{\,M}_{\odot}$ contributes to the SMBH mass budget, and is thus implicitly accounted for in the SMBH tally. For simplicity, we assume a universal lower cutoff for the stellar mass, independent of metallicity, such that $M_{\rm i, min}=0.1 \mathrm{\,M}_{\odot}$ and $M_{\rm i, max}=260 \mathrm{\,M}_{\odot}$ represent the minimum and maximum stellar mass for our idealized IMF, respectively. Integrating, we obtain the total stellar mass:
\begin{equation}\label{eq:18}
    M_{\star}=\int_{M_{\rm i, \min }}^{M_{\rm i,\max }}\left(\frac{d N}{d M_{\mathrm{i}}}\right) M_{\mathrm{i}} d M_{\mathrm{i}}
\end{equation}
in turn determining the normalizing constant. Here, we will consider both top-heavy ($\alpha = -1.35$; \citealt{Stacy2013MNRAS}) and Salpeter ($\alpha =-2.35$; \citealt{Salpeter1955ApJ...121..161S}) IMF slopes for Pop III stars, but only the Salpeter value for Pop I/II stars. Given the stellar mass function, a rough estimate for the BH mass function is:
\begin{equation}
    \left(\frac{d N}{d M_{\rm SRBH}}\right) =  \left(\frac{d N}{d M_{\mathrm{i}}}\right) \frac{d M_{\mathrm{i}}}{d M_{\rm SRBH}}.
    \label{eq:dnMBH}
\end{equation}
The derivative of the initial-to-remnant (SRBH) mass relation is evaluated by interpolating figure~2 from \cite{Husain2021MNRAS}. In the case of PISNe, for stars with an initial mass larger than $140 \mathrm{\,M}_{\odot}$ or smaller than the critical mass $M_{\star,\mathrm{c}}$, this function will return 0.

Overall, the total stellar BH mass in a halo of mass $M_{\rm h}$ at a given redshift can be estimated by integrating over the halo star formation history (SFH), as follows:
\begin{equation}
\begin{aligned}
M_{\bullet, \rm tot}\left(M_{\mathrm{h}},z\right)&=\int_{z_{\mathrm{ini}}}^{z} \mathrm{SFH}(M_{\mathrm{h}},z^{\prime})f_{\text {loss }}(t(z)-t(z^{\prime})| M_{\mathrm{i}}\geq M_{\star,\mathrm{c}}) \\ &\times f_{\mathrm{rem}} (t(z)-t(z^{\prime}))\left|\frac{dt}{dz^{\prime}}\right|\mathrm{d}z^{\prime}. \label{eq:MBH}
\end{aligned}
\vspace{-10pt}
\end{equation}
In the above expression, $f_{\text {loss }}(t| M_{\mathrm{i}}\geq M_{\star,\mathrm{c}})$ denotes the mass fraction of the stellar population contained in stars with initial masses exceeding the critical mass $M_{\star,\mathrm{c}}$ that have lifetimes shorter than $t$. This fraction is a function of the time difference between the current redshift $z$ and the redshift $z^{\prime}$ at which the stars initially form. Since there is still uncertainty regarding the onset of first star formation \citep[e.g. see fig.~2 in][]{Klessen2023ARA&A..61...65K}{}{}, for definiteness, we assume that Pop~III star formation begins at $z_{\rm ini}=50$. 

We note that the lifetime of the massive stars ($\gtrsim 10 \rm \ M_{\odot}$) capable of forming BHs is less than $\sim \mathcal{O}(10\ \mathrm{Myr})$. 
Therefore, the fraction $f_{\rm loss}(t|M_{\rm i}>M_{\star,\rm c})$ will reach the maximum at $t\gtrsim 10\ \rm Myr$. Since in most cases, the cosmic time difference between the moment of star formation ($z'$) and a given reference redshift ($z$) will be $\gg 10~ \rm Myr$, we remove the time dependence and approximate $f_{\rm loss}$ with its final maximum value for simplicity\footnote{As noted earlier, the upper mass limit for this integration depends on the stellar population. The maximum BH progenitor mass, $M_{\rm SRBH, \max }$, is 140 $\rm \ M_{\odot}$ for Pop~III stars and 260 $\rm \ M_{\odot}$ for Pop~I/II stars.}: 

\begin{equation}
\label{eq:21}
    f_{\text {loss }} \simeq  \frac{\int_{M_{\star,\mathrm{c}}}^{M_{\rm i, \max }}\left(\frac{d N}{d M_{\mathrm{i}}}\right) M_{\mathrm{i}} d M_{\mathrm{i}}}{M_{\star}}\mbox{\ .}
\end{equation}
In Equation~(\ref{eq:MBH}), $f_{\mathrm{rem}}$ signifies the fraction of the progenitor star's mass converted into (astrophysical) BH remnants, given by the ratio of the final BH mass, $M_{\mathrm {SRBH}}$, to the initial stellar mass, $M_{\mathrm{i}}$. Upon integrating over the relevant mass interval, we arrive at the remnant mass fraction, as follows:
\begin{equation}
\label{eq:22}
    f_{\text {rem }} \simeq  \frac{\int_{M_{\star,\mathrm{c}}}^{M_{\rm i, \max }}\left(\frac{d N}{d M_{\rm i}}\right) M_{\rm SRBH}(M_{\mathrm{i}}) d M_{\rm i}}{\int_{M_{\star,\mathrm{c}}}^{M_{\rm i, \max }}\left(\frac{d N}{d M_{\mathrm{i}}}\right) M_{\mathrm{i}} d M_{\mathrm{i}}}.
\end{equation}

To model the star-formation history in a halo at redshift $z$, we separately consider the formation history of Pop III and Pop I/II stars as a function of redshift, represented as $\mathrm{SFH}(M_{\mathrm{h}},z) = \mathrm{SFH}_{\rm III}(M_{\mathrm{h}},z)+\mathrm{SFH}_{\rm I/II}(M_{\mathrm{h}},z)$. In a first order approximation, we can express this as the product of the comoving volume corresponding to a given halo, and the comoving cosmic star formation rate density (SFRD):
\begin{equation}
\label{eq:23}
    \mathrm{SFH}_{\rm i}(M_{\mathrm{h}}(z),z) = \mathrm{SFRD}_{\rm i}(z)\times V_c(M_{\rm h}(z),z). 
\end{equation}
Here, $V_c(M_{\rm h}(z)) = 8\pi G M_h(z)/ 3H_0^2 $ represents the comoving volume corresponding to the halo mass $M_{\rm h}(z)$, and the subscript $i$ refers to the specific population of stars.

We note that the SFRD is an average over cosmological volumes, and is thus not accurately representing the conditions in a given halo. Star formation is dominated by Pop I/II stars at redshift $z\lesssim 10$, a phenomenon well-constrained by available data \citep[for a recent review, see][]{Klessen2023ARA&A..61...65K}. 
Accordingly, we employ the fit from the Universe Machine (Behroozi et al. 2013) for the formation history of Pop I/II stars, denoted as $\mathrm{SFH}_{\rm I/II}(M_{\mathrm{h}},z)$. The parametric fit, along with the SMBH masses, are summarized in Appendix \ref{AppendA}. However, the formation rate at $z\gtrsim 10$ remains largely uncertain due to the scarcity of available data, and the fit from the Universe Machine predicts almost no star-formation within minihaloes. As such, we use the fit for Pop~III SFRD based on a cosmological simulation presented in \cite{Liu2020MNRAS.497.2839L}, rephrased for the typical redshift $z\sim 20$ of Pop~III star formation:
\begin{equation}
    \frac{\mathrm{SFRD}_{\rm III}(z)}{\mathrm{M}_{\odot} \mathrm{yr}^{-1} \mathrm{Mpc}^{-3}}=1.1\times 10^{-5}\frac{[(1+z)/21]^{2.63}}{0.054+[(1+z)/21]^{8.55}}.
\label{eq:24}
\end{equation}

\begin{figure}
    \centering
    \includegraphics[width=0.5\textwidth]{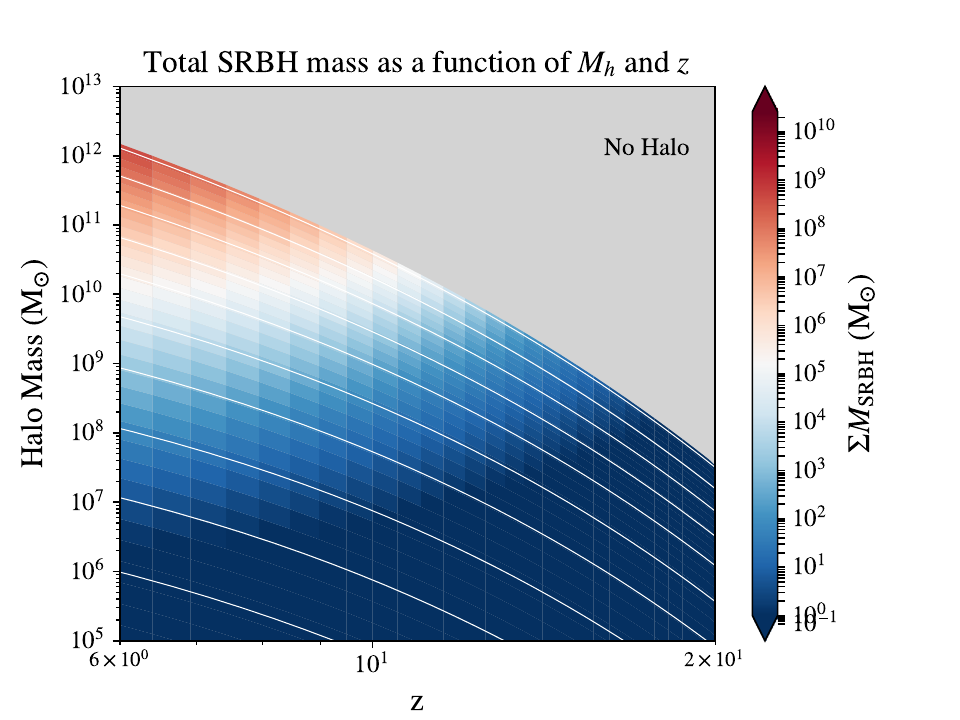}
    \vspace{-15pt}
    \caption{Cumulative BH mass from star formation throughout cosmic history. The overall star formation rate was generated by the parametric fit from the UNIVERSE\,MACHINE \citep{Zhang2023MNRAS.518.2123Z}. We explicitly add Pop~III star formation as the dominant contribution at high redshifts ($z \gtrsim 15$) \citep{Liu2020MNRAS.497.2839L}. The grey area represents the effective absence of star forming haloes,  with negligible halo number densities at those redshifts. The white lines reproduce the halo mass growth histories of select haloes.}
    \label{fig:BHMhz}
    \vspace{-10pt}
\end{figure}

Evaluating Equ. (\ref{eq:MBH}-\ref{eq:24}), we arrive at Figure~\ref{fig:BHMhz}, which summarizes the total mass of SRBHs as a function of redshift and halo mass. We observe that, generally, as the mass of the dark matter (DM) halo increases, more astrophysical black holes (SRBHs) are produced. Moreover, the total BH mass approximately scales as $\sim 10^{-4} M_{\rm h}$ at $z\simeq6$.

On the other hand, we also consider two separate fits, namely \cite{Madau2014ARA&A..52..415M} and \cite{Harikane2022ApJS..259...20H}, for the total star formation rate density (SFRD) as a first-order approximation to the star formation history (SFH) of Pop I/II stars in a galaxy using Equ.~(\ref{eq:23}). 
Recent studies have found these two fits to be reasonable bounds for the observed SFRD of galaxies at $z=9-16$ \citep{Harikane2023ApJS..265....5H}, with the work by \cite{Madau2014ARA&A..52..415M} serving as the upper bound. Therefore, at $z\gtrsim 10$, we can express the upper bound and extrapolate it to even higher redshifts as follows:
\begin{equation}
        \frac{\mathrm{SFRD}_{\rm I/II}(z)}{ \mathrm{M}_{\odot} \mathrm{yr}^{-1} \mathrm{Mpc}^{-3}}\lesssim 0.007 \frac{[(1+z)/10]^{2.7}}{0.001+[(1+z) / 10]^{5.6}}
\end{equation}

\vspace{-10pt}
\subsubsection{X-ray Binaries}

One important source of radiation feedback arises from the remnants of massive stars that form in binaries. Specifically, we focus on high-mass X-ray binaries \citep[HMXBs, reviewed by e.g.][]{Fornasini2023}, where one of the stars in the binary has an initial mass $M_{\mathrm{i}} \gtrsim M_{\star,\mathrm{c}}$. To estimate the abundance of HMXBs, we consider the cases of Pop~III and Pop~I/II stars separately. For Pop III stars, we adopt the prescription from previous studies as described in \cite{Jeon2014MNRAS.440.3778J} and \cite{Stacy2013MNRAS, Liu2021MNRAS.501..643L}, assuming that about $\sim 30 \% $ of Pop~III stars form in binaries. On the other hand, observations suggest a larger binary fraction of approximately 0.5 in the local universe \citep[e.g.][]{SanaBinaries2013A&A...550A.107S, ClarkBinaries2023MNRAS.tmp..473C}. Therefore, we apply this factor of 0.5 to Pop I/II stars. We acknowledge that not all massive binaries will evolve into HMXBs due to many factors, such as common envelope evolution and metallicity \citep[e.g.][]{Power2009MNRAS,Mirabel2011A&A...528A.149M}. Regarding those massive binary stars, for simplicity, we assume that all of them experience stages of X-ray activity and use a simple phenomenological model to calculate their radiation, thus providing an upper limit for the HMXB production and radiation from the corresponding accretion. 

Specifically, we consider a duty cycle during which HMXBs accrete mass and strongly radiate. To bracket the uncertainties, we assign a value of $0.1-1$ to represent this duty cycle, which has a weak correlation with HMXB luminosity, according to observations \citep[e.g.][]{SidoliHMXB2018MNRAS.481.2779S}. Furthermore, we assign a lifetime of $t_{\rm acc}$$\sim$ $1-10 \rm \,Myr$ to all HMXBs to allow them to accrete a substantial amount of mass ($\sim \mathcal{O}(\rm M_{\odot})$) from their companions  \citep{Mirabel2011A&A...528A.149M}. We estimate the total mass of active HMXBs in a given halo of mass $M_h$ as follows:
\begin{equation} \label{eq:HXMB}
\begin{aligned}
M_{\rm HMXB}&\left(M_{\mathrm{h}},t(z)\right)/ f_{\rm HMXB}\\=&\int_{t(z)-t_{\rm acc}}^{t(z)} \mathrm{SFH}(M_{\mathrm{h}},t') f_{\text {loss }}f_{\mathrm{rem}}\mathrm{d} t^{\prime}\\
 \approx&\ \mathrm{SFH}(M_{\mathrm{h}},t(z)) f_{\text {loss }}f_{\mathrm{rem}}t_{\rm acc}\ .
\end{aligned}
\vspace{-10pt}
\end{equation}
Here, $f_{\rm HMXB}$ refers to the aforementioned fraction of massive stars that form in binaries. The second equality is valid since $t_{\rm acc}$ is much smaller than the cosmic time $t$. We note that a comprehensive assessment of the energy feedback from HMXBs would require a dedicated study with binary population synthesis \citep[see e.g.][]{Fragos2013xrb,Fragos2013,Sartorio2023,Liu2024}, beyond the scope of the current paper.

\begin{figure}
    \centering
    \includegraphics[width=0.5\textwidth]{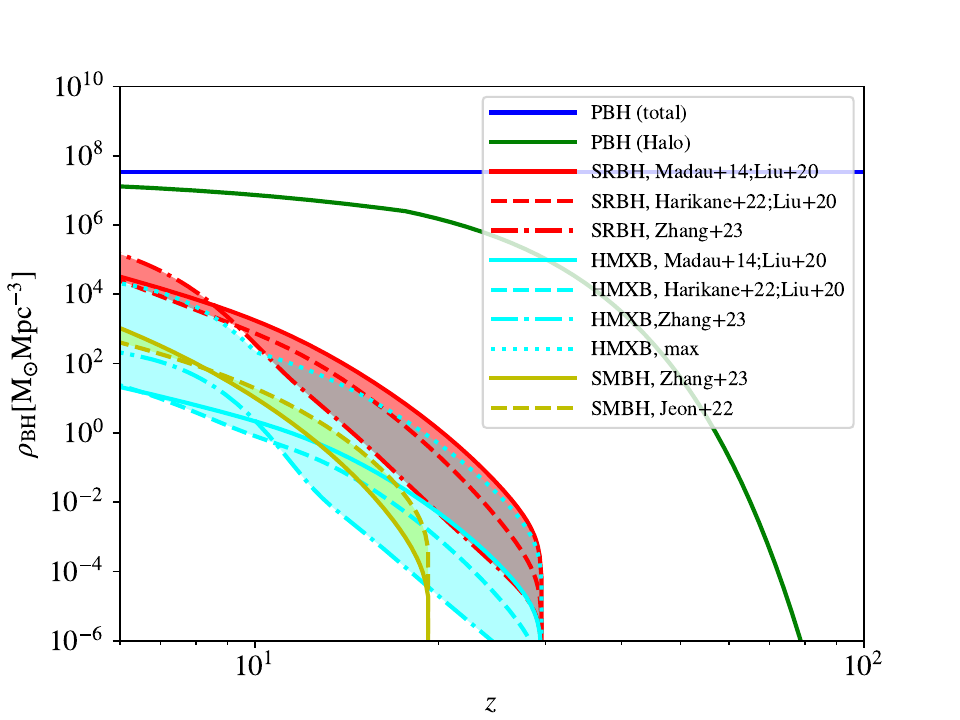}
    \vspace{-5pt}
    \caption{ The evolution of BH density from various sources: all PBHs (blue), PBHs in haloes (green), SRBHs (red), HMXBs (blue), and SMBHs (yellow). Here, we show results for a monochromatic PBH mass distribution with $M_{\rm c} = 10\mathrm{\,M}_{\odot}$, and a mass fraction in DM of $f_{\rm PBH} = 10^{-3}$. For SRBHs, we assume instantaneous formation from the remnants of massive stars. To comprehensively calculate the BH mass budget for SRBH and HMXBs, we employ the star formation history of both Pop~III and Pop~I/II stars \citep{Madau2014ARA&A..52..415M,Liu2020MNRAS.497.2839L,Harikane2022ApJS..259...20H,Harikane2023ApJS..265....5H, Zhang2023MNRAS.518.2123Z}, assuming a Salpeter IMF for Pop~I/II stars, and a top-heavy IMF for Pop~III stars. We show the maximum possible HMXB contribution (cyan dotted line), representing the limiting case of full duty cycle and accretion time of $\sim 10\rm \,Myr$. For the other cases, we assign a duty cycle of 0.1 and an accretion time of $\sim 1\rm \,Myr$. For the case of SMBHs, we use the models from \citet{Jeon2022MNRAS} (dashed) and \citet{Zhang2023MNRAS.518.2123Z} (solid). Here, Mass densities are given in comoving units.}
    \label{fig:RhoBH}
\vspace{-10pt}
\end{figure}

With the prescriptions above and integrating over the halo mass function, we can derive the average BH density for different species and redshifts. An example is shown in Figure~\ref{fig:RhoBH} for a monochromatic PBH mass distribution with $M_{\rm c} = 10\mathrm{\,M}_{\odot}$, and a mass fraction in DM of $f_{\rm PBH} = 10^{-3}$. We calculated the total mass density of PBHs to be $\sim 10^8 ~\rm M_{\odot} Mpc^{-3} $. We conclude that PBHs will dominate the BH mass budget throughout the redshift range considered here, unless $f_{\rm PBH}\lesssim 10^{-4}$. However, as more stars form and die, around $z \sim 6$, the mass density in SRBHs (including HMXBs) becomes non-negligible in dark matter haloes, at about $\sim 10^{-2}$ times the mass density of the PBHs. Moreover, we notice that the SRBH fraction in the form of active X-ray binaries is comparable to the abundance of SMBHs, both reaching a mass density level of around $\sim 10^{3} ~\rm M_{\odot} Mpc^{-3}$, generally consistent with the result from \cite{Jeon2022MNRAS}.

\section{Accretion Physics} \label{sec:accret}

In this section, we briefly describe the BH accretion physics as we aim to study the corresponding energy feedback. We consider BHs of different origins, located either in the intergalactic medium (IGM) (Section \ref{subsec:IGM}) or in haloes (Section \ref{subsec:halo}).

In general, regarding the accretion rate of a BH, we adopt a modified version of the Bondi-Hoyle steady accretion formula \citep{Tremmel2017MNRAS.470.1121T}:
\begin{equation}\label{eq:27}
    	\dot{M}_{\bullet}=\frac{4 \pi (G M_{\bullet})^{2}\rho_{\mathrm{gas}}}{v_{\rm eff}^{3}}=\frac{4\pi (G M_{\bullet})^{2}\rho_{\mathrm{gas}}}{(c_{s}^{2}+v_{\mathrm{rel}}^{2})^{3/2}}.
\end{equation}
Here, $\rho_{\mathrm{gas}} = \mu n m_{\mathrm{p}}$ represents the average gas density of the environment, given by the product of the average molecular weight $\mu = 1.22$, the gas number density $n$, and the proton mass $m_{\mathrm{p}}$. Furthermore, $c_{\mathrm{s}}$ represents the sound speed determined by the gas temperature, and $v_{\mathrm{rel}}$ is the relative velocity of the BH with respect to the gas. Thus, the effective velocity is given by $v_{\mathrm{eff}} = \sqrt{c_{\mathrm{s}}^{2}+v_{\mathrm{rel}}^{2}}$. We also note that for binary BHs the accretion rate could be modified compared to single BHs with the same mass. However, the change in the magnitude of the accretion rate is still under debate, as some simulations identify suppressed accretion rates because of the gravitational torque exerted by binaries \citep[e.g.][]{MacFadyen2008ApJ...672...83M,Miranda2017MNRAS.466.1170M}{}{}, whereas others find that binary accretion rates are not significantly modified \citep[e.g.][]{Farris2014ApJ...783..134F,Shi_2015}{}{}. Therefore, for simplicity, we assume a single BH accretion rate for all BHs.

Next, we introduce a dimensionless accretion factor:
\begin{equation}\label{eq:28}
    \quad \dot{m} \equiv \frac{\dot{M}_{\bullet}}{\dot{M}_{\mathrm{Edd}}},
\end{equation}
where $\dot{M}_{\mathrm{Edd}}$ represents the Eddington accretion rate given by:
\begin{equation} \label{eq:Medd}
    \dot{M}_{\mathrm{Edd}}=2.7 \times 10^{-7} \mathrm{M}_{\odot} \mathrm{yr}^{-1}\left(\frac{M_{\bullet}}{10 \mathrm{\,M}_{\odot}}\right)\left(\frac{\epsilon_0}{0.1}\right)^{-1}.
\end{equation}
Here, $\epsilon_0$ refers to the radiative efficiency. 

\subsection{Black Holes in the IGM}\label{subsec:IGM}
Since there is no process that allows for star formation in the IGM, we here only consider the case of PBH accretion. For the convenience of the reader, we provide a brief summary of the relevant analytical formalism, as developed in previous studies. For PBH accretion in the IGM, we adopt the self-similar model that incorporates spherical Bondi accretion with the damping effect of Compton drag and Hubble flow at high redshifts \citep{Ricotti2007ApJI,Ricotti2008ApJII,Ali-Haimoud2017PhRvD}.
For isolated PBH accretion in the IGM, the accretion factor, normalized to typical values, can be expressed as follows\footnote{We will also consider the effect of enhanced accretion by DM clothing surrounding individual PBHs, since we here assume that PBHs do not comprise the entirety of DM. This is presented in Appendix \ref{AppendC}.}:

\begin{equation}\label{eq:30}
    \dot{m}=3.40 \times 10^{-3} \lambda \left(\frac{1+z}{1000}\right)^{3}\left(\frac{M_{\bullet}}{10 \mathrm{\,M}_{\odot}}\right)\left(\frac{v_{\rm eff}}{10 \mathrm{~km} \mathrm{~s}^{-1}}\right)^{-3}.
\end{equation}
Here, $\lambda$ represents a scaling constant that represents the damping effect, numerically fitted as follows:
\begin{equation}\label{eq:31}
    \lambda=\exp \left(\frac{9 / 2}{3+\hat{\beta}^{0.75}}\right) \frac{-1+(1+\hat{\beta})^{1 / 2}}{\hat{\beta}} .
\end{equation}
The effective viscosity $\hat{\beta}$ is in turn determined by the Compton drag and Hubble flow, given by:
\begin{equation}\label{eq:32}
\begin{aligned}
\hat{\beta}=&4.4\times 10^{-4} \left(\frac{M_{\bullet}}{10\mathrm{\,M}_{\odot}}\right)\left(\frac{1+z}{1000}\right)^{3 / 2}\left(\frac{v_{\mathrm{eff}}}{10 \mathrm{~km} \mathrm{~s}^{-1}}\right) \\
& \times\left[1+56.3\left(\frac{x_{\rm e}}{10^{-3}}\right)\left(\frac{1+z}{1000}\right)^{5 / 2}\right].
\end{aligned}
\end{equation}
Here, $x_{\rm e}$ represents the ionization fraction as a function of redshift. For simplicity, we approximate $x_{\rm e}$ as a step function, in line with the CMB constraint from \cite{Plank2020A&A...641A...6P}. For $6 \lesssim z\lesssim 1000$, which extends to epochs long after recombination, we take its value to be a constant\footnote{For the case of PBH accretion in the IGM, previous studies have calculated the ionization fraction, e.g. \cite{Ziparo2022MNRAS}. In their study, although $x_{\rm e}$ changes by an order of $\sim 100$ with respect to $f_{\rm PBH}\sim 10^{-4}-1$, this factor has a negligible effect on $\hat{\beta}$.}, specifically $x_{\rm e}\sim 10^{-3}$. For $z \lesssim 6$, after reionization, this factor is set to 1. In all cases considered, we have $\hat{\beta}\ll 1$. Then, the scaling constant can be approximated as $\lambda\simeq \exp(3/2)/2 \simeq 2.24$. 

For the gas effective velocity $v_{\mathrm{eff}}$ in the IGM, we adopt the analytical approximation from \cite{Ricotti2008ApJII} and \cite{Ali-Haimoud2017PhRvD} by averaging over the Maxwell velocity distribution for PBHs:
\begin{equation}
\left\langle v_{\mathrm{eff}}\right\rangle \sim c_s\left(\frac{16}{\sqrt{2 \pi}} \mathcal{M}^3\right)^{\frac{1}{6}} \theta(\mathcal{M}-1)+c_s\left(1+\mathcal{M}^2\right)^{\frac{1}{2}} \theta(1-\mathcal{M}).
\end{equation}
Here, $\mathcal{M} = \left\langle v_{\mathrm{rel}}\right\rangle / c_{\rm s}$ represents the Mach number, and $ \left\langle v_{\mathrm{rel}}\right\rangle$ is the root-mean-square of the relative velocity between dark matter (DM) and baryons, approximated as\footnote{ We use a slightly different prescription for $ \left\langle v_{\mathrm{rel}}\right\rangle$ vs. $z$, as compared to \cite{Ricotti2008ApJII}, by considering the relative motion between DM and baryons in non-linear perturbation theory \citep[see e.g.][]{Tseliakhovich2010PhRvD..82h3520T,Stacy2011ApJ...730L...1S_StreamingDM,Dvorkin2014PhRvD..89b3519D}.}:
\begin{equation}
    \left\langle v_{\rm rel}\right\rangle \approx \min \left[1, z / 10^3\right] \times 30 \mathrm{~km}\mathrm{\,s}^{-1}.
\end{equation}
The gas sound speed, $c_{\rm s}$, is calculated from the CMB temperature, which is a function of redshift. We use the parametric fit given in \cite{DeLuca2020JCAP...04..052D}:
\begin{equation}
    c_s \simeq 5.70\left(\frac{1+z}{1000}\right)^{1 / 2}\left[\left(\frac{1+z_{\mathrm{dec}}}{1+z}\right)^\beta+1\right]^{-1 / 2 \beta} \mathrm{km} \mathrm{\,s}^{-1}.
\end{equation}
Here, $\beta = 1.72$ is a fitting parameter, and $z_{\mathrm{dec}}\simeq 130 $ the decoupling redshift for baryons from the radiation field. Combining Equ. (\ref{eq:Medd}) and (\ref{eq:30}), we can approximate the mass growth rate as:
\begin{equation}\label{eq:36}
    \dot{M}_{\bullet} \sim 2.7\times 10^{-7} \dot{m}\  \left(\frac{M_{\bullet}(t)}{10 \mathrm{\,M}_{\odot}}\right)\left(\frac{\epsilon_0}{0.1}\right)^{-1} \mathrm{M}_{\odot} \ \mathrm{yr}^{-1} .
\end{equation}
Using the above equations to calculate the accretion rate and integrating over cosmic time, we find that for PBH masses of interest, $M_{\rm PBH}\lesssim 100 \rm \ M_{\odot}$, the change in mass is $\delta M_{\rm PBH} /M_{\rm PBH} \ll 1$, which is negligible. Therefore, to make conservative estimates and simplify our calculations, for PBH accretion in the IGM, we assume a mass function $\psi(M)$ that is independent of redshift. However, we also note that, although yet to be observed, the accretion rate of PBHs, $\dot{m}$, could exceed the Eddington limit, if DM haloes are seeded by more massive PBHs (with $\gg 100 \rm\  M_{\odot}$) in isolation\, \citep[e.g. ][]{Ricotti2007ApJI,Mack2007ApJ,Ricotti2008ApJII,DeLuca2020prd}{}{}. In this scenario, the PBH can accrete a significant amount of mass by $z\lesssim 6$, thus significantly changing the shape of the PBH mass function. To fully address this effect, a dedicated simulation study would be required following up on the work by \cite{Liu2022ApJ...937L..30L}.  
\vspace{-10pt}

\subsection{Black Holes in haloes}\label{subsec:halo}

For the case of accretion within virialized haloes, we encounter more complexity compared to the previous case, since there will be BHs of various origins within a given DM halo. To simplify the calculations, we 
assume that both SRBHs and PBHs undergo Bondi accretion:
\begin{equation}\label{eq:37}
\begin{aligned}
\dot{m}= 3.2 \times 10^{-6}\left(\frac{M_{_{\bullet }}}{10 \mathrm{\,M}_{\odot}}\right)\left(\frac{n(r)}{1 \mathrm{\,cm}^{-3}}\right)\left(\frac{\mu}{1.22}\right)\left(\frac{v_{\rm eff}}{10 \mathrm{\,km} / \mathrm{s}}\right)^{-3}\mbox{.}
\end{aligned}
\end{equation}
Here, for a BH at a given radius $r$ relative to the center of the halo, $n(r)$ is the local gas number density. For the case of monochromatic PBHs, we use the simulated results from \citet{Boyuan2022MNRAS.514.2376L}, suggesting a nearly isothermal gas density profile\footnote{Here we assume that gas is always able to collapse into a quasi-isothermal sphere despite the feedback from accreting PBHs. However, collapse of gas may be delayed/halted by BH feedback in low-mass ($\lesssim 10^{8}\ \rm M_{\odot}$) haloes without atomic cooling, which can lead to shallower gas density profiles and lower BH accretion rates \citep{Liu2024}. The effects of local BH feedback on gas properties are still in debate. Our results based on the quasi-isothermal gas density profile should be regarded as optimistic estimates.}: $n(r)\propto r^{-2.2}$. Additionally, we assume the same scaling for the PBH number density. For SRBHs, we assume a similar density profile for simplicity, since the formation of SRBHs should approximately track the gas density profile, which in turn has a similar shape as the DM density profile in this regime, according to previous simulation results \citep[see fig.~7 in][]{Boyuan2022MNRAS.514.2376L}{}{}. Further simulation studies will be needed to more realistically track the location of SRBHs vs. PBHs in a halo, including any dynamical segregation effects at later stages. Meanwhile, we use the virial velocity as the effective velocity $v_{\rm eff}$ between BHs and the gas:
\begin{equation}
     v_{\rm eff} \sim \sqrt{\frac{G M_{\mathrm{h}}}{R_{\mathrm{vir}}}} \sim 5.4 \mathrm{~km} \mathrm{~s}^{-1}\left(\frac{M_{\mathrm{h}}}{10^6 \mathrm{\,M}_{\odot}}\right)^{\frac{1}{3}}\left(\frac{1+z}{21}\right)^{-\frac{1}{2}}.
\end{equation}
Here, $R_{\rm vir}$ is the virial radius of the halo, which is related to halo mass and redshift by:
\begin{equation}\label{eq:39}
R_{\mathrm{vir}} \sim 0.147\left( \frac{\Delta_{\rm c}}{18 \pi^2}\right)^{-1 / 3} \left(\frac{M_{\rm h}}{10^6 \mathrm{M}_{\odot}}\right)^{1 / 3} \left(\frac{1+z}{21 }\right)^{-1} \mathrm{kpc}, 
\end{equation}
where $\Delta_{\rm c}$ is the characteristic overdensity of the halo with respect to the cosmic average, and we adopt a fiducial value of 200. We note that for certain cases of BH accretion, Equ.~(\ref{eq:37}) could yield rates larger than the Eddington limit. Conservatively, we enforce $\dot{m}\lesssim 1$ throughout, and thus do not consider any super-Eddington accretion regime, which we cannot reliably represent with our idealized modeling here.
%

The dynamics and density distribution of PBHs with an extended mass function can be complex, and its effect on SRBH and SMBH formation would require future simulations. Additionally, SRBHs will also have a non-trivial mass function, adding further complexity to the situation. For now, we can derive order of magnitude estimations. We assume that the extended PBH mass distribution will not significantly change the density profiles of both gas and BHs in the host halo. To calculate the gas number density at different radii, we normalize to the value at the virial radius:
\begin{equation}\label{eq:40}
\begin{aligned}
    n (R_{\rm vir})& \sim  \Omega_{\mathrm{b}}(1+z)^3 \frac{3 H_0^2}{8 \pi G}\frac{\Delta_{\rm c}}{\mu m_p} \\ &\sim 0.33 \left(\frac{\Delta_{\rm c}}{18 \pi^2}\right)\left(\frac{\mu}{1.22}\right)^{-1}\left(\frac{1+z}{21 }\right)^{3} \rm cm^{-3}.\\
\end{aligned}
\end{equation}

 Similarly, since no previous simulations have considered the co-evolution of PBHs and SRBHs, for simplicity, we apply the same formula to calculate the accretion rate for both BH classes. To further simplify our calculations, we impose additional assumptions, as follows. For more massive SRBHs (with progenitor stellar mass $\gtrsim 260 \rm \ M_{\odot}$), we assume that they gradually sink towards the halo center to become seeds for SMBHs, or get captured by the pre-existing central SMBH. Thus, we cut off the progenitor stellar mass functions for SRBHs at $\gtrsim 260 \rm \ M_{\odot}$ and restrict their radii to be inside of $0.001 R_{\rm vir}$. Typically, for a $10^9 \ \mathrm{M}_{\odot}$ halo at $z\simeq 10$, the virial radius is $R_{\rm vir} \sim 3\ \rm kpc $.  The inner region, $r<0.001 R_{\rm vir}$, thus has a size $\sim \mathcal{O}(\rm pc)$, and can only contain a SMBH. For PBHs following a log-normal mass function, we numerically integrate from $0.1-10 M_{\rm c}$ to cover the plausible mass range.
 
 Furthermore, we assume a time independent mass function for both PBHs and SRBHs (excluding HMXBs). 
We find that any evolution of the BH mass function in a halo is likely small, as we argue below. Since the halo will also capture PBHs during its growth, we can compare the total PBH accretion rate with the capture rate of PBHs by DM halo accretion. A fiducial value for the capture rate is the accretion rate calculated for the Millennium simulation \citep{McBride2009MNRAS.398.1858M}, modified for a contribution from PBH DM, as expressed in the PBH mass fraction $f_{\rm PBH}$:
\begin{equation}
\begin{aligned}
\langle\dot{M}_{\rm PBH, tot}\rangle \sim & 0.02 {\rm \ M_{\odot} \mathrm{yr}^{-1}}\left(\frac{M_{\rm h}}{10^{12} \rm \ M_{\odot}}\right)^{1.094} \left(\frac{f_{\rm PBH}}{10^{-3}}\right)\\
& \times(1+1.75 z) \sqrt{\Omega_m(1+z)^3+(1-\Omega_{m})}\mbox{\ .}
\end{aligned}
\end{equation}
This is of the same order of magnitude as the accretion rate obtained by integrating over all PBHs with a characteristic mass of $100\rm \ M_{\odot}$ in a $10^{12} M_{\rm h}$ halo at $z=0$, as given by Equ.(\ref{eq:36})-(\ref{eq:40}). Therefore, at higher redshifts, we would expect a much larger PBH capture rate than the total accretion rate. For SRBHs, we find that the increase in total SRBH mass is proportional to the star formation rate, i.e. $\sim 0.01 \ \mathrm{SFH}(M_{\rm h},z)$, as calculated using Equ.(\ref{eq:MBH}), which is also significantly larger than the total accretion rate. In both cases, the newly increased BH population will likely preserve the mass function at formation. In general, however, the overall BH mass function in a given halo will result from the competition between the BHs that already sit in the halo versus newly accreted or born BHs. Although we do not explore such cases here, a more complete treatment should broaden the discussion.
A more detailed simulation study of the mass function evolution within a DM halo would be necessary in future work. For now, we can accommodate only minor changes to the overall mass function of SRBHs and PBHs. Finally, for HMXBs and SMBHs, we note that feedback could affect the accretion efficiency in complex ways.\footnote{ Feedback effects in SMBHs play a dual role, where kinetic-mode feedback self-regulates gas accretion, decreasing the Eddington ratios and limiting mass growth \citep[see e.g.][]{Weinberger2018MNRAS.479.4056W}{}{}, while super-Eddington accretion, in the absence of feedback, can significantly increase SMBH mass in a fraction of the Eddington time \citep[e.g.][]{Li2012MNRAS.424.1461L}{}{}.
 In HMXBs, the accretion efficiency is significantly impacted by the presence of accretion disks, leading to super-Eddington rates in ultraluminous X-ray sources (ULXs), and modulated by interactions between stellar winds and the black hole's gravitational influence, as seen in systems like Cyg X-1 \citep[for a review on HMXBs see][]{Fornasini2023}{}{}.
} For simplicity and consistent with the prescription in \cite{Jeon2014MNRAS.440.3778J}, we impose Eddington-limited accretion throughout ($\dot{m}=1$).

\vspace{-10pt}

\section{Radiative signature} 

\label{sec:radfed}
\subsection{Cosmic Radiation Background}
In this section, we aim to study the contribution from all BHs, as summarized in Figure~\ref{fig:evoscheme}, to the cosmic radiation background. Here, we focus on cases with monochromatic PBH mass distribution, and defer fully explored results for an extended mass distribution to future work. We begin by splitting the energy density $u$ from BH feedback into two parts based on the BH habitat:
\begin{equation}
u=u_{\mathrm{IGM}}+u_{\mathrm{halo}},
\end{equation}
where $u_{\rm IGM}$ represents the energy feedback from PBHs accreting in the IGM, and $u_{\rm halo}$ that from all BHs in DM haloes. Based on the DM halo mass, we further decompose the feedback in collapsed structures into contributions from SRBHs (including HMXBs), PBHs, and SMBHs, as follows:

\begin{equation}\label{eq:42}
    u_{\rm halo}=\begin{cases} u_{\rm PBH},& \\ \hspace{20ex}  (M_{\min}(z) \leq M_{\rm h}<M_{\rm H_2-cool})&\\  u_{\rm SRBH}+u_{\rm HMXB}+u_{\rm PBH}, & \\  \hspace{20ex} (M_{\rm H_2-cool} \leq M_{\rm h}<10^8 \mathrm{M}_{\odot})  &\\u_{\rm SRBH}+u_{\rm HMXB}+u_{\rm PBH}+u_{\rm SMBH}.&\\    \hspace{20ex} (10^8 \mathrm{M}_{\odot}  \leq M_{\rm h}< M_{\max}(z)) & \end{cases}
\end{equation}
Here, several DM halo mass limits are introduced. Specifically, $M_{\min}(z)$ refers to the smallest halo mass that could host a PBH, given by $M_{\min} = \max (M_c \Omega_{\rm m}/f_{\rm PBH}\Omega_{\rm DM} , M_{\rm vir})$, where $M_{\rm vir}$ refers to the smallest mass that can virialize at a given redshift into a halo structure with a virial temperature $\gtrsim 100 \mathrm{\,K}$. We also notice some other pertinent mass scales, such as $\sim M_{\rm c} (1+z_{\rm eq})/(1+z)$, describing a halo seeded by a single PBH \citep[see e.g.][]{Ricotti2007ApJI,Ricotti2008ApJII}. This effect will further increase the radiation output since more minihaloes are present at high redshift. In our modeling, we derive a conservative estimate with the current choice of mass threshold and address the PBH-seeded halo case in Appendix~\ref{AppendC}. We defer a more complete understanding of halo accretion at high redshifts to future study with simulations.

$M_{\rm H_2-cool}$ is the smallest mass of a DM halo capable of forming Pop~III stars via H$_2$ cooling, and it is fitted by $M_{{\rm H}_2-\mathrm{cool}} \simeq 1.54 \times 10^5 \rm \ M_{\odot}\left[(1+z)/31\right]^{-2.074}$ \citep{Trenti2009ApJ...694..879T}. $M_{\rm max}(z)$ is the maximum allowed mass of DM haloes at redshift $z$, given by the progenitor of $\sim 10^{16} \rm \ M_{\odot}$ haloes at $z=0$, following their growth history\footnote{For a typical value, $M_{\rm max}\lesssim 10^{13}\rm \ M_{\odot}$ at redshift 6.}. Haloes with mass larger than $M_{\rm max}(z)$ are excluded due to low number density. The growth trajectories of all haloes are modeled by \cite{Behroozi2013ApJ...770...57B} and \cite{Behroozi2018MNRAS.477.5382B} (see Appendix \ref{AppendA}). For PBHs in haloes, since this halo growth model might not be applicable at high redshift, we set a uniform upper limit of $10^{13} \rm \ M_{\odot}$ for numeric approximation.

To calculate each term, we need to revisit the accretion physics and study the radiative efficiency based on the accretion rate. As a good estimation, we adopt the sub-grid model from \cite{Negri2017MNRAS_subgrid} to capture the transition from a radiatively efficient thin disk to an optically thick and advection-dominated accretion flow (ADAF) case. The relationship between BH accretion rate $\dot{m}$ and bolometric luminosity is given by:
\begin{equation}\label{eq:44}
    L=\epsilon_{\mathrm{EM}} \dot{M}_{\bullet} c^{2},\ \epsilon_{\mathrm{EM}}=\frac{\epsilon_{0} A \dot{m}}{1+A \dot{m}}.
\end{equation}
where $\epsilon_{\mathrm{EM}}$ is the electromagnetic efficiency, $A=100$, and $\epsilon_{0} = 0.057 $ the radiative efficiency for optically thin accretion. Note that $\epsilon_{\mathrm{EM}}$ reduces to $\epsilon_{0}$ for higher accretion rates, $\dot{m} \gg 0.01$.

To evaluate the total luminosity from BH accretion in a halo of mass $M_{\rm h}$ at redshift $z$, we sum over the mass distributions of BHs, and normalize with respect to the BH mass function. In a general form, the total luminosity from BHs of a single origin is given by:
\begin{equation}\label{eq:45}
\begin{aligned}
L_{\bullet}(M_{\rm h},z) &= \sum_{i} \Delta N_{\bullet,\mathrm{i}}(M_{\bullet},M_{\rm h}) L(M_{\bullet}) \\ &= M_{\bullet, \mathrm{tot}} (M_{\rm h}) 
\frac{\int_{M_{\bullet, \min }}^{M_{\bullet, \max }} \psi(M_{\bullet}) L(M_{\bullet})dM_{\bullet}}{\int_{M_{\bullet, \min }}^{M_{\bullet, \max }} \psi(M_{\bullet}) M_{\bullet} d M_{\bullet}},
\end{aligned}
\end{equation}
where $\Delta N_{\bullet,\mathrm{i}}(M_{\bullet},M_{\rm h})$ is the number of BHs in mass bin $i$ of width $\Delta M_{\bullet}$, within a halo of mass $M_{\rm h}$ at redshift $z$. Furthermore, $L(M_{\bullet})$ is the luminosity associated with BHs of mass $M_{\bullet}$, and $\psi(M_{\bullet})$ is the BH mass distribution function in a general form. To obtain the total number of BHs, we divide the total BH mass in a halo, $ M_{\bullet \mathrm{,tot}} (M_{\rm h})$, by the average BH mass, as calculated in the denominator of the second equality above. When integrating, $M_{\bullet, \min }$ and $M_{\bullet, \max }$ represent the respective BH mass limits.
For PBHs, we substitute the denominator with the characteristic BH mass $M_{\rm c}$, and integrate over the mass range of $0.1-10 M_{\rm c}$. For SRBHs and HMXBs, we employ Equ.~(\ref{eq:IMF}-\ref{eq:dnMBH}) with the mass limits discussed in Section~\ref{sec:ABHs}. For SMBHs, we assume that each massive halo contains one central SMBH, and that the luminosity is given by the Eddington value, $L_{\bullet}\simeq L_{\rm Edd}(M_{\bullet})\simeq \epsilon_{0}\dot{M}_{\rm Edd}c^{2}$.

Once we have the aggregate power from one type of BH in a halo of mass $M_{\rm h}$ at redshift $z$, the (proper) average power density, $ j_{\bullet}(z)$, is obtained by integrating over the halo mass function, as mentioned previously:
\begin{equation}
    j_{\bullet}(z) = \int_{M_{\rm min}(z)}^{M_{\rm max}(z)}(1+z)^{3} n\left(M_{\rm h},z\right) L_{\bullet}(M_{\rm h},z) dM_{\rm h},
    \label{eq:46}
\end{equation}
where $n\left(M_{\rm h},z\right)$ is the comoving number density of haloes per unit halo mass, given by Equ.(\ref{eq:HMF}). The factor of $(1+z)^{3}$ is applied to derive the physical number density at a given $z$. We note that, if the maximum halo mass $M_{\rm max}$ is smaller than the lower limit $M_{\rm min}$ at some redshift $z$, there is no contribution from that particular type of BH. The lower integration limit $M_{\rm min}$ depends on the BH type. For PBHs located in a halo, $M_{\rm min} =\max (M_c \Omega_{\rm m}/f_{\rm PBH}\Omega_{\rm DM} , M_{\rm vir})$. For SRBHs, we choose $M_{\rm min} =M_{\rm H_2-cool}$, whereas for SMBHs, we adopt $M_{\rm min}=10^8 \mathrm{\,M}_{\odot}$ as the minimum halo mass that can host an AGN.
The proper bolometric power density $u$ at redshift $z$ from BHs can be obtained by integrating the power density at a given redshift $z^\prime$ from different sources over cosmic time:
\begin{equation}\label{eq:energy density}
    u(z) =    \int_{z_{\rm i}}^{z}\left(\frac{1+z}{1+z^\prime}\right)^{4}j_{\bullet}(z^\prime) \left| \frac{dt}{dz^\prime}\right| dz^\prime\mbox{\ ,}
\end{equation}
where $z_{\rm i}$ and $z$ are the initial and final redshift of consideration, respectively. Here, we consider the most distant PBH source emitting at $z_{\rm i} \simeq 1000$. For the case of SRBHs, we start the integration from $z_{\rm i} \simeq 50$.

\begin{figure}
\centering
\begin{subfigure}{0.45\textwidth}
    \caption{PBH mass fraction: $f_{\rm PBH} = 10^{-3}$}
    \includegraphics[width=\textwidth]{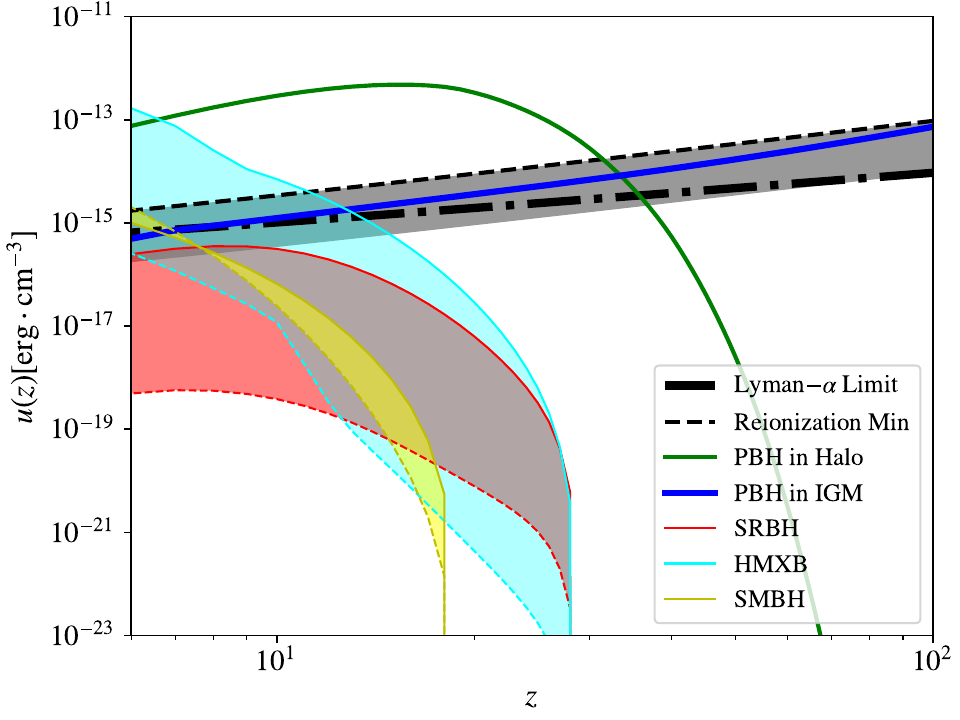}
\end{subfigure}
\hfill
\begin{subfigure}{0.45\textwidth}
    \caption{PBH mass fraction: $f_{\rm PBH} = 10^{-5}$}
    \includegraphics[width=\textwidth]{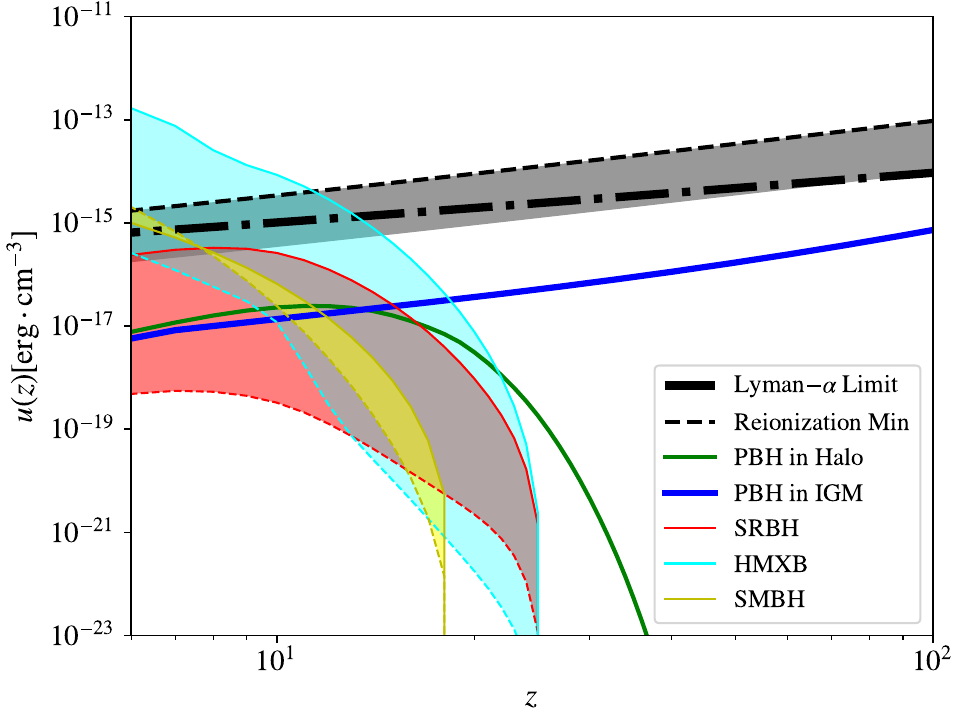}
\end{subfigure}
\caption{Integrated (bolometric) energy density from BH accretion feedback of different origins vs. redshift: PBHs in the IGM (blue), in haloes (green), SRBHs (red), HMXBs (cyan), and SMBHs (yellow). For PBHs, we assume a monochromatic mass distribution with $M_{\rm c} = 10 \rm \ M_{\odot}$, and mass fractions of $f_{\rm PBH} = 10^{-3}$ (upper panel) and $10^{-5}$ (lower panel). Here, we consider the same range for the abundance of BHs as shown in Figure~\ref{fig:RhoBH}, when deriving possible radiation outputs. For comparison, we also plot the minimum Lyman-$\alpha$ flux required for Wouthuysen-Field coupling (dot-dashed line; from \citealt{CiardiMadau2003ApJ...596....1C}), as well as the minimum UV flux required for reionization (dashed line and grey-shaded area; from \citealt{Madau1999ApJ...514..648M}).}
\label{fig:Ltot}
\end{figure}

The results are shown in Figure~\ref{fig:Ltot} for different PBH mass fractions: $f_{\rm PBH} = 10^{-3}$ (upper panel) and $f_{\rm PBH} = 10^{-5}$ (lower panel), both assuming a monochromatic mass distribution with $M_{\rm c} = 10 \mathrm{\,M}_{\odot}$. The figure confirms our previous statement that PBHs can only make up a small fraction of dark matter, as otherwise reionization would have occurred too early (see below for further discussion). If we assume a PBH mass fraction of $10^{-3}$, the feedback power from PBHs in haloes starts to dominate at $z \lesssim 30$. We also note that, since SMBHs and HMXBs are accreting at a much higher rate than SRBHs, their emission power at lower redshift ($z\lesssim 10$) starts to surpass that from SRBHs. In addition, we discuss the energy output from the enhanced accretion rate within PBH-seeded halos in Appendix \ref{AppendC}.

For context, we compare our predicted bolometric luminosities to the lower limits on the cosmic radiation density required for key cosmological signatures. To render the 21-cm hyperfine structure line of neutral hydrogen detectable in the form of absorption, its internal spin temperature needs to be coupled to the IGM kinetic temperature. An effective way to accomplish this coupling is through the Wouthuysen-Field effect \citep{Wouthuysen1952AJ.....57R..31W, Field1959ApJ...129..536F}, where Lyman-$\alpha$ photons scatter with H atoms, thus indirectly modifying the hyperfine level populations  \citep{Madau1997ApJ...475..429M}. We can estimate the required Lyman-$\alpha$ energy density, in proper units and assuming a flat spectrum, as follows: $u_{\rm Ly \alpha} \simeq  4\pi J_{\alpha} \nu_{\alpha}/c \gtrsim 3.5\times 10^{-14}[(1+z)/7]\ \rm erg \,cm^{-3} $ \citep{CiardiMadau2003ApJ...596....1C}. Similarly, we can estimate the minimum energy density in H-ionizing UV radiation required to reionize the IGM: $u_{\rm UV} \simeq  4\pi J_{\rm UV} \nu_{\rm UV}/c \gtrsim 2 \times 10^{-15} [(1+z)/7]^{1.5}\ \rm erg \,cm^{-3} $\citep{Madau1999ApJ...514..648M}. However, to gauge the possible impact of PBH DM on cosmic reionization, we need to go beyond bolometric quantities, towards a more careful spectral modelling for accreting BHs (see below). Furthermore, any reionization constraints have to take into account the escape fraction of UV radiation from the source host haloes.  

To obtain the (angle-averaged) radiation intensity at frequency $\nu$ and a given redshift $z$, we need to calculate the specific emission coefficient, $j_{\bullet, \nu^\prime}(z^\prime)$, from sources at higher redshifts $z^\prime>z$ with $\nu^\prime=\nu (1+z^\prime)/(1+z)$. Applying the same procedure as for Equ. (\ref{eq:45}-\ref{eq:46}) but replacing the bolometric luminosity with the specific luminosity, i.e. $L\rightarrow L_{\nu}$ (per BH), $L_{\bullet}\rightarrow L_{\bullet,\nu}$ (per halo), and $j_{\bullet}\rightarrow j_{\bullet,\nu}$, we first sum over the contributions from all BHs in a halo, and then integrate the halo mass function to obtain the specific emission coefficient. Here, the features of the emerging spectrum from a single BH depend mostly on the normalized accretion rate $\dot{m}$, determining whether accretion operates in the thin-disk or ADAF regime. We employ the prescription of \cite{Takhistov2022JCAP}, as summarized in Appendix~\ref{AppendB}, to calculate the specific luminosity $L_{\nu}(M_{\bullet})$ for any BH. Finally, integrating all contributions from preceding redshifts, the resulting specific intensity $J_{\nu}$ at $z$ is \citep[e.g.][]{Schauer2019ApJ...877L...5S}:
\begin{equation}
    J_{\nu}(z)=\frac{c}{4\pi}\int_{z_{\rm i}}^{z}\left(\frac{1+z}{1+z^\prime}\right)^{3}j_{\bullet, \nu^\prime}(z^\prime)\left|\frac{d t}{d z^\prime}\right| d z^\prime.
\label{eq:48}
\end{equation}
We note that both Equ. (\ref{eq:energy density}) and (\ref{eq:48}) have no absorption term, thus assuming a transparent IGM and maximizing the radiation level from BH sources. Therefore, $z_{\rm i}$ denotes the largest redshift that a given BH source starts to emit light. To calculate the intensity for a given frequency band, we simply integrate $J_{\nu}$ over the corresponding band, and $z_{\rm i}$ denotes the highest redshift below which radiation produced by BHs can reach $z$ in the given band. In the following sub-sections, we will specifically discuss the intensity in the UV and X-ray bands 
, comparing them with observational constraints. We emphasize that when evaluating the resulting feedback on the IGM, the escape fraction $f_{\rm esc}$ of radiation from the host haloes should be taken into account, depending on the waveband considered. For now, and unless otherwise noted, we set the escape fraction to 1, thus maximizing the radiative output from the accreting BH sources (further discussed below).

\vspace{-10pt}

\subsection{UV Intensity}

\begin{figure}

    \centering
    \includegraphics[width=0.5 \textwidth]{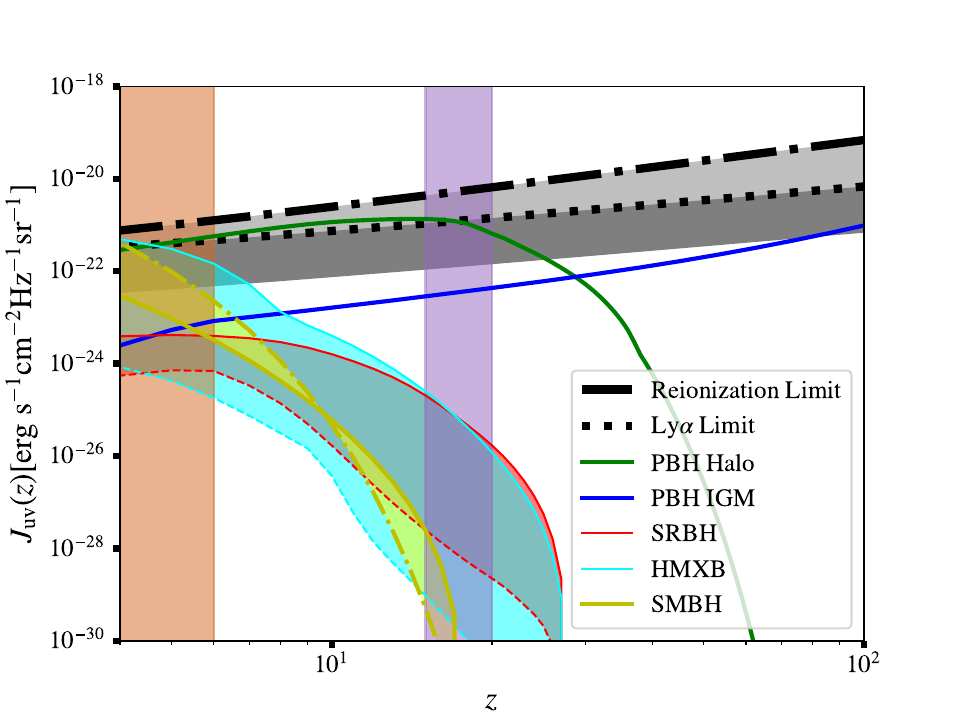}
    \caption{Background intensity at the Lyman limit, $\nu_{\rm UV}={\rm 13.6\,eV}/h$, vs. redshift. Here, we again consider various BH sources (PBHs, SRBHs, HMXBs, and SMBHs), differentiating between halo and IGM PBH contributions. For the sake of simplicity, we assume a monochromatic mass function for PBHs with a characteristic mass of $10 \ \rm M_{\odot}$ and a mass fraction of $10^{-3}$. We show the minimum intensity, $J_{\rm UV}$, required for reionization \citep{Madau1999ApJ...514..648M}, as well as the minimum Lyman-$\alpha$ background, $J_{\alpha}$, needed for efficient Wouthuysen-Field (WF) coupling \citep{CiardiMadau2003ApJ...596....1C}. For the comparison to the latter, we for simplicity assume $\nu_{\alpha}\simeq \nu_{\rm UV}$, and for both limits, we indicate uncertainties with a factor-of-ten range (grey shaded areas). We also indicate the timing constraint for effective WF coupling at $z\sim 15-20$ (purple shaded area), inferred from the EDGES 21cm-cosmology measurement \citep{EDGES2019MNRAS.483.1980M}. Additionally, the brown region denotes the reionization limit at $z\sim 6$. }
    \label{fig:Juv}
\end{figure}

In Figure~\ref{fig:Juv}, we present our calculations of the (average) UV intensity from BHs as a function of redshift, assuming a monochromatic mass function for PBHs with a characteristic mass $M_{\rm c}=10\ \rm M_{\odot}$ and a mass fraction $f_{\rm PBH}=10^{-3}$. We evaluate the radiation intensity at $\nu_{\rm UV}=13.6~\mathrm{eV}/h$, and assume for simplicity the same value at $\nu_{\alpha}$, i.e. $J_{\alpha}\simeq J_{\rm UV}$, which is sufficiently precise for our purpose here. It is evident that while the contribution from PBHs in the IGM initially dominates, it becomes significantly smaller compared to the radiation from BHs in haloes for $z\lesssim 30$. Our estimation of the UV power from SMBHs is about an order of magnitude lower than the findings in \cite{Jeon2022MNRAS}. This discrepancy arises because of differences in the spectral modeling, employing a more detailed approach here (see Appendix~\ref{AppendB}). Another contributing factor might be our model's prediction of a BH mass that is smaller by roughly a factor of $\sim 10$ when compared to observed high-redshift quasars \citep[e.g.][]{Decarli2018ApJ...854...97D,Wang2021ApJ...907L...1W}.

According to a confluence of observations, the IGM was fully reionized by $z\sim 6$ (reviewed in \citealt{Robertson2010}), with early evidence based on the Gunn-Peterson absorption trough in high-$z$ quasar spectra \citep[e.g.][]{Becker2001AJ....122.2850B}. When juxtaposing the reionization limit with the radiation background from all BHs, it is clear that PBHs consistently dominate over other BH sources prior to reionization, as long as $f_{\rm PBH}\gtrsim 10^{-3}$. For our fiducial value of cosmic PBH density (0.1\% of all DM), we can see that at $z\lesssim 15$, the UV intensity from PBHs (within haloes) becomes comparable to the level required for reionization, suggesting that they could be a significant source of ionizing photons for the neutral hydrogen in the IGM. Taken at face value, such large UV intensities would allow for reionization much earlier than the canonical redshift of $z\sim 6$. However, when calculating the resulting UV intensity, we have assumed that all ionizing photons produced via BH accretion inside a halo can escape from the host ($f_{\rm esc} = 1$). More realistic treatments, despite the inherent uncertainty, have constrained this factor to be significantly smaller, around 0.1 for typical host systems in the pre-reionization Universe \citep[e.g.][]{Gnedin2008ApJ...672..765G,Khaire2016MNRAS.457.4051K}. Given our ballpark calculations, as presented in Figure~\ref{fig:Juv}, PBH DM could contribute significantly to the sources of reionization. Depending on the (uncertain) UV escape fraction, this yields another constraint on the PBH DM fraction, roughly giving $f_{\rm PBH}\lesssim 10^{-3}/f_{\rm esc}$.

Similarly, we assess whether the PBH contribution to the cosmic Lyman-$\alpha$ background can induce Wouthuysen–Field coupling, thus imprinting a global absorption signature in the redshifted 21-cm radiation \citep[e.g.][]{Pritchard2012}. Such a signature was reported by the Experiment to Detect the Global Epoch of Reionization Signature (EDGES), suggesting that the spin temperature of neutral hydrogen was coupled via the Wouthuysen-Field effect to the kinetic gas temperature around $z\simeq 15 - 20$ \citep{Bowman2018Natur.555...67B,EDGES2019MNRAS.483.1980M}. This claimed detection remains a topic of vigorous debate, with more recent spectral radiometer measurements that have found no such signal \citep{Singh2022}. Prior research \citep{Mena2019PhysRevD.100.043540} modeled the alterations in the kinetic temperature of IGM gas due to PBH heating, constraining the PBH abundance to $f_{\rm PBH}\lesssim 10^{-3}$ for $M_{\rm c} = 10\rm \ M_{\odot}$. Using the same characteristic mass, we find a similar constraint on the PBH abundance, as $f_{\rm PBH}> 10^{-3}$ would imply effective Wouthuysen-Field coupling at redshifts larger than the EDGES range (see Fig.~\ref{fig:Juv}). We note that in arriving at this conclusion, we have assumed an escape fraction for Lyman-$\alpha$ photons of close to unity, which is plausibe for the dust-poor galaxies at cosmic dawn \citep[e.g.][]{Jaacks2018, Smith2019}. As our results indicate, accreting PBHs located within haloes are possibly potent sources for Lyman-$\alpha$ background radiation at high redshifts, whereas the other BH sources considered here fall short in producing a significant number of Lyman-$\alpha$ photons. For a discussion of the contribution of stellar sources to the cosmic Lyman-$\alpha$ radiation field, we refer to the extensive literature on this topic \citep[e.g.][]{Fialkov2014rich,Schauer2019ApJ...877L...5S,Gessey-Jones2022}.

\subsection{X-ray Intensity}

\begin{figure}
\centering   \centering
\begin{subfigure}{0.5\textwidth}
    \includegraphics[width=\textwidth]{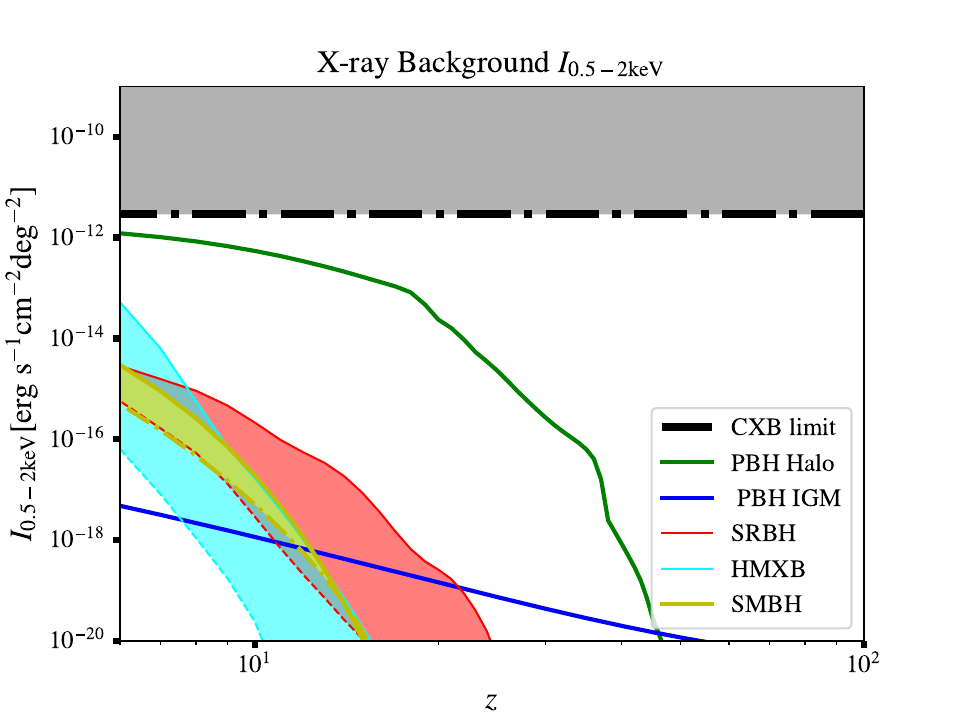}
\end{subfigure}
\hfill
\begin{subfigure}{0.5\textwidth}
    \includegraphics[width=\textwidth]{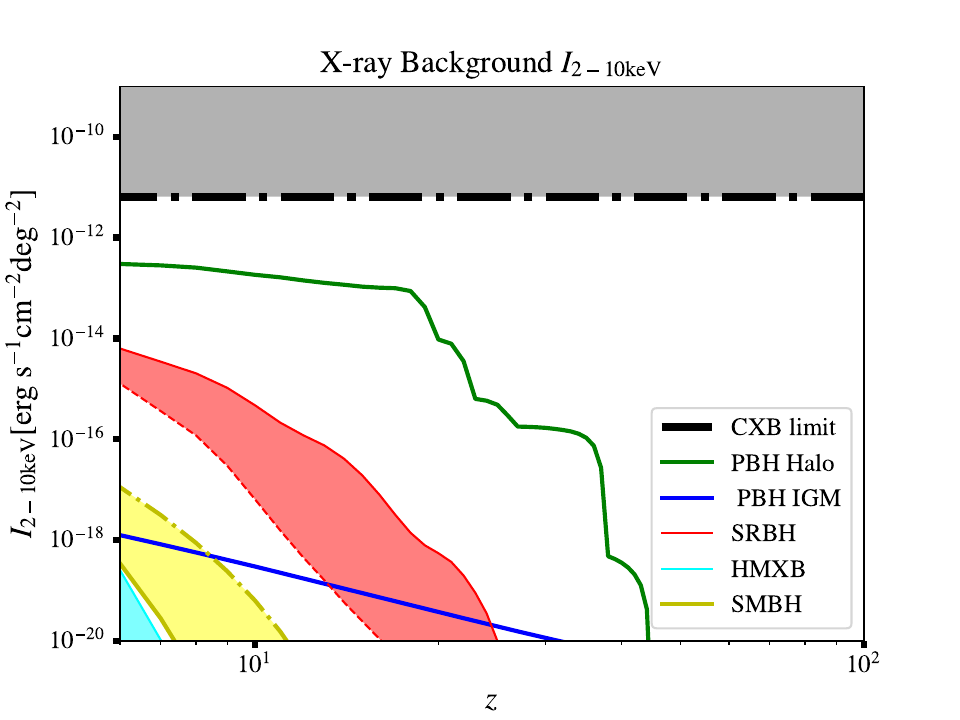}
\end{subfigure}    
\caption{Contribution of accreting high-$z$ BH sources to the present-day ($z=0$) X-ray background. X-ray intensities are plotted against redshift for all BH sources considered here, applying the same set of parameters as used in Figure~\ref{fig:Juv}. The black dash-dotted line represents the unresolved X-ray background \citep{Cappelluti2017ApJ...837...19C,Cappelluti2022ApJ}. The upper panel shows the soft X-ray band with a photon energy in the $[0.5-2 \rm keV]$ range, while the lower panel depicts the integrated hard X-ray band with photon energies of $[2-10\rm keV]$.}
\label{fig:X-ray}
\end{figure}

In Figure \ref{fig:X-ray}, we illustrate the contribution of BHs from various sources to the unresolved portions of the present-day hard ($[2-10~ \mathrm{keV}]$) and soft ($[0.5-2~\mathrm{keV}]$) X-ray backgrounds, again considering PBHs with $M_{\rm c}=10 \ \rm M_{\odot}$ for a monochromatic mass function and $f_{\rm PBH}=10^{-3}$. 
The X-ray background intensity in a given energy band $[E_{1}-E_{2}]$ is $I_{\rm E_{1}-E_{2}}=\int _{\nu_{1}}^{\nu_{2}}J_{\nu}d\nu$, $\nu_{i}=E_{i}/h$, $i=1,2$, where $h$ is the Planck constant.
We emphasize that any unresolved sources are preferentially expected at $z\gtrsim 6$, given the low luminosity of typical high-redshift systems. It is thus natural to consider accreting BHs at cosmic dawn when accounting for the unresolved CXB. Note that when assessing the contribution of sources at $z^\prime$ to the radiation observed at $z=0$, we use $\nu^\prime = (1+z^\prime)\nu$ for the integration involved in Equ.~(\ref{eq:48}). 

Our results for PBHs agree with the findings of \cite{Ziparo2022MNRAS}, when using identical parameter choices. However, by incorporating the effects of DM-baryon streaming and Compton drag, which diminish the accretion rate, we determine that PBHs in the IGM contribute only a minuscule fraction to the overall X-ray radiation. For $z\lesssim 30$, the BH feedback power from haloes becomes dominant, similar to the behaviour in the (rest-frame) UV band. Comparing our predicted radiation levels to the unresolved CXB inferred from deep Chandra observations \citep{Cappelluti2017ApJ...837...19C,Cappelluti2022ApJ}, we conclude that the accretion feedback from PBHs, located within haloes, may account for a substantial part of the missing X-ray flux. More specifically, in the soft X-ray band ($[0.5-2 ~\rm keV]$), radiation from PBHs at $z\gtrsim 6$ almost reaches the intensity level of the unresolved X-ray background, whereas in the hard X-ray band ($[2-10 ~\rm keV]$), radiation from these PBHs approaches the unresolved background intensity, but falls somewhat short. In both bands, the radiation from PBHs dominate over all other BH sources by at least an order of magnitude. 

We note that neutral H and He in the IGM absorb soft X-ray photons ($[0.5-2~\mathrm{keV}]$) more efficiently than hard X-ray radiation ($[2-10~\mathrm{keV}]$), due to the approximately $\nu^{-3}$ dependence of the photo-ionization cross sections \citep[e.g.][]{Wilms2000ApJ...542..914W}. As a result, the upper panel in Figure~\ref{fig:X-ray} represents the upper limit for the soft X-ray contribution from PBHs. In summary, our calculations suggest that radiation from accreting PBHs at high redshifts may play a significant role in accounting for the unresolved portion of the cosmic X-ray background, given a PBH abundance parameter of $f_{\rm PBH}\gtrsim 10^{-3}$. 

\section{Summary and Conclusions}\label{sec:summary}

We have investigated the impact of PBHs as a dark matter component on structure formation and radiation backgrounds in the early Universe. Taking into account the presence of PBHs, we have analyzed the energy feedback resulting from the accretion of matter by all BHs of various origins across cosmic history. Our study has led us to the following key findings and conclusions:
\begin{itemize}
    \item[(i)]From the constructed PBH-$\Lambda$CDM Universe, we find that even in the absence of initial clustering of PBHs, their presence will accelerate the formation of virialized structures (haloes), as the Universe becomes clumpy with the emergence of minihaloes of mass $\lesssim 10^6\rm \ M_{\odot}$. Increasing both PBH mass abundance and characteristic mass parameters will accelerate structure formation, with the former being more significant, as shown in Figure~\ref{fig:fcol}. 
    \item[(ii)]In the IGM, only PBHs are present. However, within DM haloes, we need to consider BHs formed through astrophysical channels in addition to PBHs. As the first stars begin to form, the remnants of these massive stars give rise to SRBHs. However, even assuming that PBHs constitute only a small fraction of the DM ($f_{\rm PBH}\lesssim 10^{-3}$), we find that HMXBs and SMBHs start to generate comparable accretion power as PBHs within halo environments only at lower redshifts $z\lesssim 7$ (see Figs.~\ref{fig:RhoBH} and \ref{fig:Ltot}).
    \item[(iii)] Considering the evolving bolometric radiation energy density produced by BH accretion, we have observed that the power density from PBH accretion in DM haloes begins to dominate over that from accretion in the IGM at $z\lesssim 30$. Initially, the radiative output from PBHs dominates, but it becomes comparable to that from SRBHs around $z\sim 7$. However, if the PBH abundance were much greater than $f_{\rm PBH}\sim 10^{-3}/f_{\rm esc}$, as shown in Figs.~\ref{fig:Ltot} and~\ref{fig:Juv}, the Universe would undergo reionization too early, in violation of existing constraints. This further supports the conclusion that PBHs within the solar mass range cannot be the primary component of dark matter.
   \item[(iv)] We have found that by changing the mass function of PBHs from a monochromatic distribution to a log-normal one, there is a minor change in the total energy density, by a factor of approximately 2. Moreover, with a larger mass dispersion parameter $\sigma$, the overall energy density further increases. This can be explained by the fact that 
   the scaling of the accretion power $L$ with respect to BH mass is non-linear with a power-law index $>2$, 
   as shown in Equ. (\ref{eq:28}-\ref{eq:Medd}), (\ref{eq:37}) and (\ref{eq:44}). However, since the combination of characteristic mass and PBH abundance parameters is highly constrained by \cite{Kuhnel2017PhRvD}, any effect from changing the mass function from monochromatic to log-normal is likely small.
\item[(v)] By modeling the spectral energy distribution (SED) of BH accretion feedback (see Appendix~\ref{AppendB}), we can calculate the energy feedback at specific frequencies (bands). From Figure~\ref{fig:Juv}, we conclude that accretion onto PBHs could be an important source for generating sufficient UV and Lyman-$\alpha$ background radiation to affect cosmic reionization or to produce detectable absorption in the (redshifted) 21-cm line via efficient Wouthuysen-Field coupling. However, for $f_{\rm PBH}\lesssim 10^{-3}$, no existing constraints would be violated. On the other hand, when comparing the X-ray power from accreting BHs with the limits on the unresolved X-ray background, as shown in Figure~\ref{fig:X-ray}, we find that the total power from BH sources in DM haloes is comparable to the present-day unresolved X-ray background limit. Again, the accretion power from PBHs in the IGM is negligible. For PBHs within haloes, the X-ray power generated is always dominant over that by other BHs, in both soft and hard X-ray bands, with the difference in the latter being even stronger. The PBH contribution may thus reach the intensity level of the unresolved soft X-ray background, but falls somewhat short of the level of the hard X-ray background, for plausible values of $f_{\rm PBH}$. We note that any conclusions for the soft X-ray background depend on the uncertain modeling of the optical depth to photo-ionization in the neutral IGM.
\end{itemize}
We would like to acknowledge several limitations in our study. Firstly, concerning the mass function of PBHs, we note that the predicted accretion rates in the IGM and virialized halos are negligibly small (see Equ.~\ref{eq:37}), rendering the mass function effectively time independent, within the characteristic mass range of $M_{\rm c} \sim 1-100 ~ \mathrm{M}_{\odot}$. Previous studies \citep[e.g.,][]{Hasinger2020JCAP,DeLuca2020prd,Yuan2023arXiv230309391Y} have attempted to address this issue using a semi-analytical approach with accretion models from \cite{Mack2007ApJ,Ricotti2008ApJII}, where the accretion rate could be significantly enhanced within haloes seeded by PBHs in the IGM environment. We note that in the case of extended or monochromatic PBH mass functions with a large characteristic mass $M_{\rm c} \gg 10 ~ \mathrm{M}_{\odot}$, the time-independent assumption will likely no longer be a good approximation. For PBHs in a virialized halo environment, changes to the mass function become more complex. Here, additional factors need to be considered, such as the distribution within haloes of PBHs with an extended mass function, the case of infalling PBHs through halo accretion from the cosmic web or through mergers with other haloes, and the merger of PBHs with other BHs. Selected recent studies have considered the possibility that BH binaries or massive BH seeds could form though PBH captures and mergers~\citep[see e.g.][]{Hayasaki2016PASJ...68...66H, Ali-Haimoud2017PhysRevDMerger,DeLuca2020JCAP...04..052D,DeLuca2023PhRvL.130q1401D}{}{}. 
In this work, for the sake of simplicity, we ignored the effect of gravitational capture or merger, and assumed a time-independent PBH mass function, for both IGM and halo environments throughout cosmic history.
 Therefore, a more accurate modeling of the change in the mass function can be achieved through cosmological simulations, which we leave as a topic for future studies.

Any predictions for the radiation feedback from all BHs are subject to the broader uncertainties in modeling BH number densities and halo gas density profile (see footnote 9). As an example, not all galaxies will host a central SMBH, giving rise to AGN feedback. Furthermore, for HMXBs we may have overestimated the accretion rate and its duty cycle. Additionally, we have not considered the scattering/absorption attenuation caused by hydrogen and helium atoms due to the challenges in calculating source distributions. However, taking this additional factor into account would not significantly alter our main conclusions since it equally applies to BHs in haloes. Overall, we can still conclude that distinguishing PBHs from other BHs solely based on the cosmic radiation background is challenging.

In future work, we plan to test the robustness of our modeling, such as  how halo density profiles may be modified in the presence of BHs, with cosmological simulations. Additionally, we will consider the PBH-generated emissions in other bands, in particular the near-infrared and radio bands. Such studies will serve to provide a more complete interpretation for ongoing or upcoming sky surveys, such as with Euclid \citep{Euclid2018LRR....21....2A} or the Square Kilometre Array (SKA) \citep{SKAWeltman2020PASA...37....2W}, to detect the imprints left on the early IGM  by the presence of PBH dark matter.

\section*{Acknowledgements}
BL is supported by the Royal Society University Research Fellowship.

\section*{Data Availability}

The data and codes used in this article will be shared upon reasonable request to the corresponding authors.



\bibliographystyle{mnras}
\bibliography{Main} 




\appendix

\section{Halo Fitting Formulae}\label{AppendA}

The total stellar mass, as a function of redshift and halo mass, is provided by \cite{Behroozi2013ApJ...770...57B,Behroozi2019MNRAS.488.3143B,Behroozi2020MNRAS.499.5702B,Zhang2023MNRAS.518.2123Z}, based on a parametric fit to observational data. In this section, we outline how we utilize their model to calculate the total SRBH and SMBH mass within a halo.

Initially, we apply the parametric fit for a single halo accretion history, following the model detailed in the appendix of \citet{Behroozi2013ApJ...770...57B}. This model is initially based on the growth trajectory of the progenitor of a $10^{13} \rm \ M_{\odot}$ halo, $M_{13}(z)$, and scaled to the haloes of different masses with the parametric fit. For a halo of mass $M_0$ (in units of $\rm \ M_{\odot}$) at redshift $z=0$, its progenitor mass $M_h$ at redshift $z$ is given by:
\begin{equation}
    \begin{aligned}
M_h\left(M_0, z\right) & =M_{13}(z) 10^{f\left(M_0, z\right)},\\
M_{13}(z) & =10^{13.276} \rm M_{\odot}(1+z)^{3.00}\left(1+\frac{z}{2}\right)^{-6.11} \exp (-0.503 z), \\
f\left(M_0, z\right) & =\log _{10}\left(\frac{M_0}{M_{13}(0)}\right) \frac{g\left(M_0, 1\right)}{g\left(M_0, \frac{1}{1+z}\right)}, \\
g\left(M_0, z\right) & =1+\exp \left(-4.651\left( \frac{1}{1+z}-a_0\left(M_0\right)\right)\right) , \\
a_0\left(M_0\right) & =0.205-\log _{10}\left[\left(\frac{10^{9.649} \rm M_{\odot}}{M_0}\right)^{0.18}+1\right].
\end{aligned}
\end{equation}

For the remainder of the parametric model, we adopt the fitted parameters from \cite{Zhang2023MNRAS.518.2123Z} to ensure consistency. The star formation rate (SFR, in units of $\mathrm{\, M}_{\odot} \mathrm{yr}^{-1}$) as a function of halo mass and redshift is fitted by a double power law
\begin{equation}
    \begin{aligned}
\mathrm{SFR}_{\mathrm{SF}}\left(M_h, a, V_h, \epsilon_h, \alpha_h,\beta_h\right) & =\frac{\epsilon_h}{v_h^{\alpha_h}+v_h^{\beta_h}}\rm M_{\odot} \mathrm{yr}^{-1}, \\
v_h & =\frac{v_{\mathrm{Mpeak}}}{V_h \cdot \mathrm{km} \mathrm{s}^{-1}},
\end{aligned}
\end{equation}
where $v_{\mathrm{Mpeak}}$ is the maximum velocity within a halo of mass $M_h$ at redshift $z$:

\begin{equation}
    \begin{aligned}
v_{\mathrm{Mpeak}}\left(M_h, a\right) & =200 \mathrm{~km} \mathrm{~s}^{-1}\left[\frac{M_h}{M_{200 \mathrm{kms}}(a)}\right]^{1 / 3}, \\
M_{200 \mathrm{kms}}(a) & =\frac{1.64 \times 10^{12} \mathrm{M}_{\odot}}{\left(\frac{a}{0.378}\right)^{-0.142}+\left(\frac{a}{0.378}\right)^{-1.79}}.
\end{aligned}
\end{equation}

Given that we have also assumed a halo growth history, this $M_h$ will change with respect to the redshift and eventually evolve into a halo of mass $M_0 $ at $z=0$. The remaining parameters ($V_h, \epsilon_h, \alpha_h,\beta_h$) can be fitted with the following values:

\begin{equation}
\begin{aligned}
\log _{10}(V_h)= & 2.289+1.548(a-1)+1.218 \ln (1+z)-0.087 z,\\
\log _{10}(\epsilon_h)= & 0.556-0.944(a-1)-0.042\ln (1+z)+0.418 z,\\
\alpha_h= & -3.907+32.223(a-1) +20.241 \ln (1+z)-2.193 z,\\
\beta_h= & 0.329+2.342(a-1)+0.492 z. \\
\end{aligned}
\end{equation}

Here, $a = 1/(1+z)$ is the scale factor.
\begin{figure}
    \centering
    \includegraphics[width=0.5\textwidth]{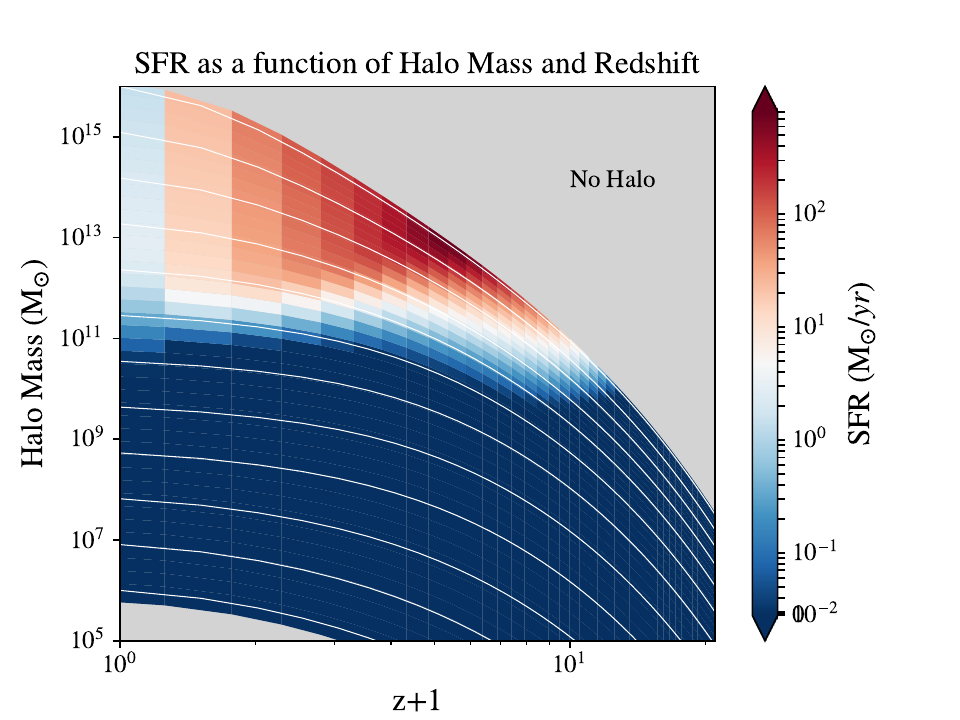}
    \vspace{-15pt}
    \caption{Star formation rate (SFR), in units of $\rm \ M_{\odot} \mathrm{\,yr}^{-1}$, as a function of redshift and host halo mass, from the parametric fit generated by the UNIVERSEMACHINE \citep{Zhang2023MNRAS.518.2123Z}. Here, in addition to the overall star formation rate, we add a Pop~III component as the dominant star formation process at high redshifts ($z \gtrsim 13$)\citep{Liu2020MNRAS.497.2839L}. Similar to Figure~\ref{fig:BHMhz}, the grey region represents the part of parameter space for the growth trajectories of $\gtrsim 10^{16}\ \rm M_{\odot }$ haloes at $z=0$, that are not expected to exist in the observable Universe due to low number densities. White lines represent halo growth trajectories for a single halo.}
    \label{fig:SFR}
\end{figure}
The resulting SFR as a function of redshift and current halo mass at redshift $z$ is plotted as a color map in Figure~\ref{fig:SFR}. Since this parametric fit only provides the SFR at low redshift ($z=0-10$), which corresponds to the SFR of higher metallicity stars, we also incorporate the SFR for Pop~III stars from \cite{Liu2020MNRAS.497.2839L}, using Equ.~(\ref{eq:23}-\ref{eq:24}). We find that it generally aligns with the JWST observation of star-forming galaxies at redshift $z\simeq 10-16$ in terms of SFR \citep{Harikane2023ApJS..265....5H}. For higher redshifts $z\gtrsim 10$, due to the lack of data, we extrapolate the fitted star formation rate from the UNIVERSEMACHINE to approximate the stellar mass.

To calculate the SMBH mass in a halo of mass $M_{\rm h}$ at redshift $z=1/a-1$, we require a few additional steps. Firstly, we calculate the total stellar mass $M_{\star}(M_h, a)$ as 

\begin{equation}
\begin{aligned}
\log _{10}\left(M_{\star}\right) & =\log _{10}\left(\epsilon_{\star} M_1\right)+f\left(\log _{10}\left(\frac{M_h}{M_1}\right)\right)-f(0), \\
f(x) & =-\log _{10}\left(10^{\alpha_{\star} x}+1\right)+\delta_{\star} \frac{\left(\log _{10}(1+\exp (x))\right)^{\gamma_{\star}}}{1+\exp \left(10^{-x}\right)},
\end{aligned}
\end{equation}

with fitted parameters ($v_{\star},\epsilon_{\star},M_1,\alpha_{\star},\delta_{\star},\gamma_{\star}$):

\begin{equation}
\begin{aligned}
v_{\star}= & \exp \left(-4 a^2\right), \\
\log _{10}(\epsilon_{\star})= & -1.777-0.006(a-1)v_{\star}-0.119(a-1), \\
\log _{10}\left(M_1\right)= & 11.514+\left(-1.793(a-1)-0.251z\right) v_{\star}, \\
\alpha_{\star}= & -1.412+0.731(a-1) v_{\star}, \\
\delta_{\star}= & 3.508+\left(2.608(a-1)-0.043 z\right) v_{\star}, \\
\gamma_{\star}= & 0.316+\left(1.319(a-1)+0.279 z\right) v_{\star}.
\end{aligned}
\end{equation}

Once we have the stellar mass for a halo at redshift $z$, the corresponding bulge mass $M_{\text {bulge }}$ can be calculated using the following equations:
\begin{align}
M_{\text {bulge }} & =\frac{f_z(z) M_{\star}}{1+\exp \left\{-1.13\left[\log _{10}\left(M_{\star} / 10^{10.2}\rm M_{\odot}\right)\right]\right\}}, \\ f_z(z) & =\frac{z+2}{2 z+2}.
\end{align}

\begin{figure}
\centering
\includegraphics[width=0.5\textwidth]{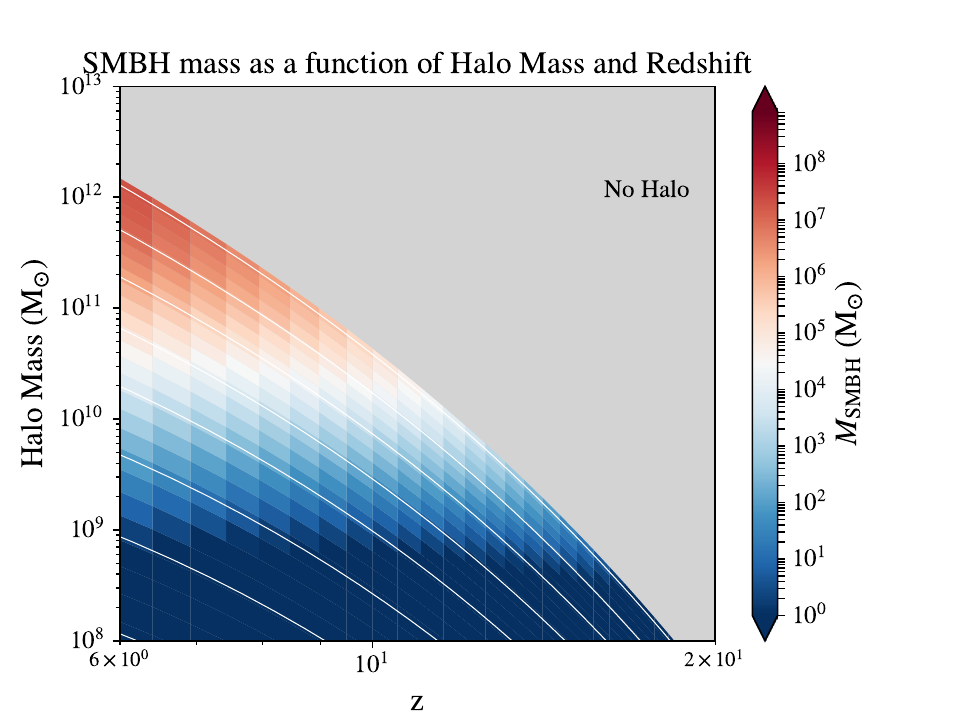}
\vspace{-15pt}
\caption{Average SMBH mass as a function of redshift and halo mass, from the parametric fit generated by the UNIVERSEMACHINE \citep{Zhang2023MNRAS.518.2123Z}. Here, we truncate the mass that could serve as SMBH host at $10^8\rm \ M_{\odot}$. Similar to Figure~\ref{fig:BHMhz}, the grey area represents the combination of parameters where no halo formation is allowed, and white lines represent individual halo growth histories.}
\label{fig:SMBH}
\end{figure}

Finally, $\tilde{M}_{\mathrm{SMBH}}\left(M_{\text {bulge }}, \gamma_{\bullet},\beta_{\bullet}\right)$, the average SMBH mass corresponding to the bulge mass $M_{\text {bulge }}$ is determined by the following fitting formula:
\begin{equation}
\begin{aligned}
& \log _{10} \tilde{M}_{\mathrm{SMBH}}=\beta_{\bullet}+\gamma_{\bullet} \log _{10}\left(\frac{M_{\text {bulge }}}{10^{11} \rm \ M_{\odot}}\right), \\
\end{aligned}
\end{equation}
where
\begin{equation}
\begin{aligned}
\gamma_{\bullet}= & 1.028+0.036(a-1)+ 0.052 z, \\
\beta_{\bullet}= & 8.343-0.173(a-1) +0.044 z.
\end{aligned}
\end{equation}
The resulting plot of the average SMBH mass in a given host halo of mass $M_h$ at redshift $z$ is given in Figure~\ref{fig:SMBH}. When compared with the SRBH mass abundance within a halo, we find that it is smaller by at least one order of magnitude in higher mass haloes at $z\gtrsim 6$. However, the higher accretion rate for SMBHs will compensate for this difference by generating more energy feedback while growing in mass.

\section{Accreting BH Spectra}\label{AppendB}

Next, we summarize the prescription for the radiation feedback spectrum from BH accretion, as described in \cite{Takhistov2022JCAP}. To streamline the analysis, we categorize the accretion process into two scenarios based on the normalized accretion factor $\dot{m}$: the thin-disk regime and the advection-dominated accretion flow (ADAF) regime. The former represents the case of disk formation from surrounding gas with sufficiently large angular momentum, whereas the latter implies quasi-spherical accretion where the angular velocity of the gas is smaller than the Keplerian velocity \citep[see, e.g.][]{Yuan2014ARA&A..52..529Y}{}{}. 

For BH accretion in the IGM and in virialized haloes, we employ a combination of the two regimes depending on the BH accretion rate\footnote{We note that in section~3.3 of \cite{Ricotti2008ApJII}, a threshold halo mass of $\lesssim 1000 M_{\odot}$ is introduced, below which the formation of an accretion disk is prevented.}, assuming a viscosity parameter $\alpha=0.1$. Conversely, for SMBHs and HMXBs accreting at the Eddington rate, we assume a thin disk regime, where the blackbody spectrum is dominant.

\textbf{Thin disk ($\dot{m}\geq 0.07 \alpha$)}

In the thin disk regime, the temperature profile of the accretion disk $T(r)$ can be modeled as a function of the radius \citep[see also][]{Pringle1981ARA&A..19..137P}{}{}:
\begin{equation}
    T(r)=T_i\left(\frac{r_i}{r}\right)^{3 / 4}\left[1-\left(\frac{r_i}{r}\right)^{\frac{1}{2}}\right]^{\frac{1}{4}}.
\end{equation}
Here, $r_i$ is the inner radius of the disk, which is equal to three times the Schwarzschild radius $r_s$. The temperature $T_i$ at the inner radius is derived from the balance between gravitational energy release and radiative cooling:
\begin{equation}
    T_{\rm i}=\left(\frac{3 G M_{\bullet} \dot{M}_{\bullet}}{8 \pi r_i^3 \sigma_{\rm SB}}\right)^{1 / 4}=1.3\times 10^3 \ \mathrm{eV}/k_{\rm B} \, \dot{m}^{\frac{1}{4}}\left(\frac{M_{\bullet}}{10 \mathrm{M}_{\odot}}\right)^{-\frac{1}{4}},
\end{equation}
where $\sigma_{\rm SB}$ is the Stefan-Boltzmann constant, and $k_{\rm B} $ the Boltzmann constant.

The maximum temperature $T_{\rm max} = T(r_{\rm max})$ of the disk is attained at $r_{\rm max}=1.36 r_i$, as given by the temperature profile. The minimum temperature $T_{\rm o} = T(r_{\rm out})$ is reached at the outer radius of the disk:
\begin{equation}
    r_{\text {out }} \simeq 1.16 \times 10^{10} r_{\rm s}\left(\frac{M_{\bullet}}{10 \mathrm{M}_{\odot}}\right)^{\frac{2}{3}}\left(\frac{v_{\rm eff}}{10 \mathrm{~km} / \mathrm{s}}\right)^{-\frac{10}{3}}.
\end{equation}
This outer radius is determined by equating the angular momentum of the gas from the surrounding environment to the angular momentum of the accretion disk. Here,$v_{\rm eff}$ is the same effective velocity as in Equ.~(\ref{eq:27}).


The overall shape of the spectrum is the result of the blackbody spectrum contributing from all parts of the accretion disk, taking the following form:

\begin{equation}
 L_\nu= \begin{cases}
    c_\alpha\left(\frac{T_{\max }}{T_o}\right)^{\frac{5}{3}}\left(\frac{h\nu}{k_{\rm B}  T_{\max }}\right)^2 & (\nu<k_{\rm B}  T_o /h )\\c_\alpha\left(\frac{h \nu}{k_{\rm B}  T_{\max }}\right)^{\frac{1}{3}} & 
(k_{\rm B} T_o /h<\nu<k_{\rm B} T_{\max }/h) \\  c_\alpha\left(\frac{h\nu}{k_{\rm B}  T_{\max }}\right)^2 e^{1-\frac{h \nu}{k_{\rm B}  T_{\max }}} & (\nu > k_{\rm B}  T_{\max }/h) \\ 
\end{cases}. \label{eq:b4}
\end{equation}

Here, $c_\alpha$ is the normalizing flux constant such that the integration over all frequencies gives a total power of $0.057 \dot{M}_{\bullet} c^2$, which recovers the subgrid model \citep{Tremmel2017MNRAS.470.1121T}.

\textbf{ADAF ($\dot{m}< 0.07 \alpha$)}:

The ADAF disk dominates when the accretion rate $\dot{m}$ is significantly sub-Eddington\footnote{For a complete derivation, see e.g. \citet{Narayan1995ApJ...452..710N}.}. In this regime, we primarily consider the spectrum from the sychrotron radiation and Inverse-Compton (IC) scattering process:
\begin{equation}
    L_{\nu}=  \begin{cases} 
        L_{\nu_p}\left(\frac{\nu}{\nu_p}\right)^{5/2} & (\text{Sychrotron},\  \nu< \nu_p) \\
        L_{\nu_p}\left(\frac{\nu}{\nu_p}\right)^{-\alpha_c}& (\text{IC},\ \nu_p \leq \nu \lesssim 3k_{\rm B}T_e / h) 
    \end{cases} .\label{eq:B5}
\end{equation}
where the peak frequency $\nu_p$ is 
\begin{equation}
\begin{aligned}
    \nu_p= & 1.83 \times 10^{3} \mathrm{eV} / h \\ &  \dot{m}^{\frac{3}{4}} \left(\frac{\alpha}{0.1}\right)^{-\frac{1}{2}}\left(\frac{1-\beta}{1 / 11}\right)^{\frac{1}{2}}\theta_e^2\left(\frac{r_i}{3 r_s}\right)^{-\frac{5}{4}}\left(\frac{M_{\bullet}}{10 \mathrm{M}_{\odot}}\right)^{-\frac{1}{2}}, \\
\end{aligned}
\end{equation}
and the corresponding peak luminosity:
\begin{equation}
    L_{\nu_p}=5.06 \times 10^{37} \frac{\mathrm{erg}}{\mathrm{s} \cdot \mathrm{eV}/h} \dot{m}^\frac{3}{2} \alpha^{-1}(1-\beta)  \theta_e^5 \left(\frac{r_i}{ r_s}\right)^{-\frac{1}{2}}\left(\frac{M_{\bullet}}{10 \mathrm{M}_{\odot}}\right).
\end{equation}
To ensure overall consistency, we normalize the product of the peak frequency and luminosity, $\nu_p L_{\nu_p}$, with the results from Equ. (\ref{eq:44}).

In these equations, $\beta$ is a partition parameter for total pressure originating from the gas and magnetic field. with values equaling $10/11$. $\theta_e$ is the dimensionless electron temperature written in units of electron rest mass $m_e$:
\begin{equation}
\begin{aligned}
    \theta_e  = & \frac{k_{\rm B} T_e}{m_e c^2}=\frac{T_e}{5.93 \times 10^9 K}\\\simeq &  0.17 A_c^{-\frac{1}{7}} \delta^{\frac{1}{7}} \alpha^{\frac{3}{14}}(1-\beta)^{-\frac{1}{14}} \left(\frac{r_i}{r_s}\right)^{\frac{3}{28}}\left(\frac{M_{\bullet}}{10 \mathrm{M}_{\odot}}\right)^{\frac{1}{14}} \dot{m}^{-\frac{5}{28}},
\end{aligned}
\end{equation}
where $A_c^{-\frac{1}{7}}$ is a pre-factor related to the relative contribution of IC to the total power depending on the accretion rate. Since we do not further divide the ADAF case, we will assume a uniform value of 1.1. $\delta$ is the electron heating efficiency approximated by a value of $0.3$.

The exponent $\alpha_c$ for the IC part of the spectrum in Equ.~(\ref{eq:B5}), is determined by the following expression:
\begin{equation}
    \alpha_c = - \frac{\ln \tau_{\rm es}}{\ln A },
\end{equation}
where $\tau_{\rm es} = 12.4 \dot{m} \alpha^{-1} (r_{i}/r_s)^{-1/2}$ is the optical depth for electron scattering, and $A = 1 +4\theta_e + 16 \theta_e^2$ is the amplification factor. Typically, for a $10 \rm \,M_{\odot}$ BH that accretes at a rate of $\dot{m} \sim 10^{-6}$, this exponent will yield a value of $\alpha_c \sim 2$, where the slope becomes asymptotically flatter with increasing accretion rate $\dot{m}$ and BH mass $M_{\bullet}$. When the accretion is efficient, reaching the critical limit for the thin disk of regime $\dot{m}\sim 10^{-2}$, for the same BH mass of $10 \rm \,M_{\odot}$, the power-law exponent approaches $\alpha_c \sim 1$.

\section{PBH Accretion Enhanced by Halo Clothing}\label{AppendC}

In this section, we summarize the changes to the accretion factor when PBHs are clothed by haloes, following the work of \citet{Ricotti2007ApJI, Mack2007ApJ, Ricotti2008ApJII,DeLuca2020prd}. We define a parameter $\kappa$ as the ratio of the Bondi radius to the radius of the clothed halo:
\begin{equation}\label{eq:c1}
\kappa = 2.2 \times 10^{-3} \left( \frac{1+z}{1000} \right) \left( \frac{M_h}{{\rm M}_\odot} \right)^{\frac{2}{3}} \left( \frac{v_{\text{eff}}}{10 \text{ km s}^{-1}} \right)^{-2}.
\end{equation}
Here, $M_h$ is the halo mass seeded by PBHs, given by $M_h \sim M_{\text{c}} (1+z_{\text{eq}})/(1+z)$. In our case, since we choose the characteristic PBH mass to be smaller than $\sim 100 {\rm\,M}_{\odot}$, we estimate $\kappa < 2$ for $z\gtrsim 30$. The original scale parameter, $\lambda$, in the relevant equations for the accretion rate, Equ.~(\ref{eq:30}) to (\ref{eq:32}), is modified as follows: 
\begin{equation}
\lambda^h \equiv \bar{\Gamma}^{\frac{2-\hat{\alpha}}{\hat{\alpha}-1}} \lambda(\hat{\beta}^h), \quad \hat{\beta}^h \equiv \kappa^{\frac{2-\hat{\alpha}}{\hat{\alpha}-1}} \hat{\beta},
\end{equation}
where $\hat{\alpha}$ is the power of the surrounding density profile, and
\begin{equation}
\bar{\Gamma} = \left(1 + 10\hat{\beta}^h\right)^{\frac{1}{10}} \exp\left(2 - \kappa\right) \left(\frac{\kappa}{2}\right)^2.
\end{equation}
Here we adopt $\hat{\alpha} = 2.2$, consistent with the density profiles in the simulations by \cite{Boyuan2022MNRAS.514.2376L} and similar to the value $\hat{\alpha}=2.25$ adopted in \citet{Ricotti2008ApJII} for the Bertschinger self-similar solution. Furthermore, we use the corrected accretion parameters $\lambda^h$  and $\hat{\beta}^h$ to replace the value of $\lambda$ and $\hat{\beta}$ in Equ.~(\ref{eq:30}) and (\ref{eq:31}).

On the other hand, at $z\lesssim 30$, we find $\kappa \gtrsim 2$, which means the DM halo behaves similarly to a point mass of $M_h$. Therefore, we will apply a correction to Equ.~(\ref{eq:30}) as:
\begin{equation}\label{eq:C4}
    \dot{m}=0.042 \lambda \left(\frac{1+z}{1000}\right) \left(\frac{M_{\bullet}}{10 \mathrm{\,M}_{\odot}}\right)\left(\frac{v_{\rm eff}}{10 \mathrm{~km} \mathrm{~s}^{-1}}\right)^{-3},
\end{equation}
using the original equations, Equ.~(\ref{eq:31}) and (\ref{eq:32}) to calculate the accretion constant $\lambda$.
To evaluate the energy density, we consider contributions from BHs residing in different environments. However, distinguishing between accretion in the IGM versus accretion in halo environments is challenging, especially at redshifts $z \gtrsim 30$. To tentatively make this distinction, we set the same mass boundary $M_{\min} = \max(M_c \Omega_{\text{m}}/f_{\text{PBH}}\Omega_{\text{DM}}, M_{\text{vir}}, M_{\text{c}} (1+z_{\text{eq}})/(1+z) )$, distinguishing virialized haloes that could host more than one PBH from those seeded by a single PBH. We treat the latter case similarly to the IGM case but calculate it using the enhanced accretion rate. More simulation work is needed in the future to better address the effects of PBH halo seeding.

\begin{figure}
\centering
\begin{subfigure}{0.45\textwidth}
    \caption{PBH mass fraction: $f_{\rm PBH} = 10^{-3}$}
    \includegraphics[width=\textwidth]{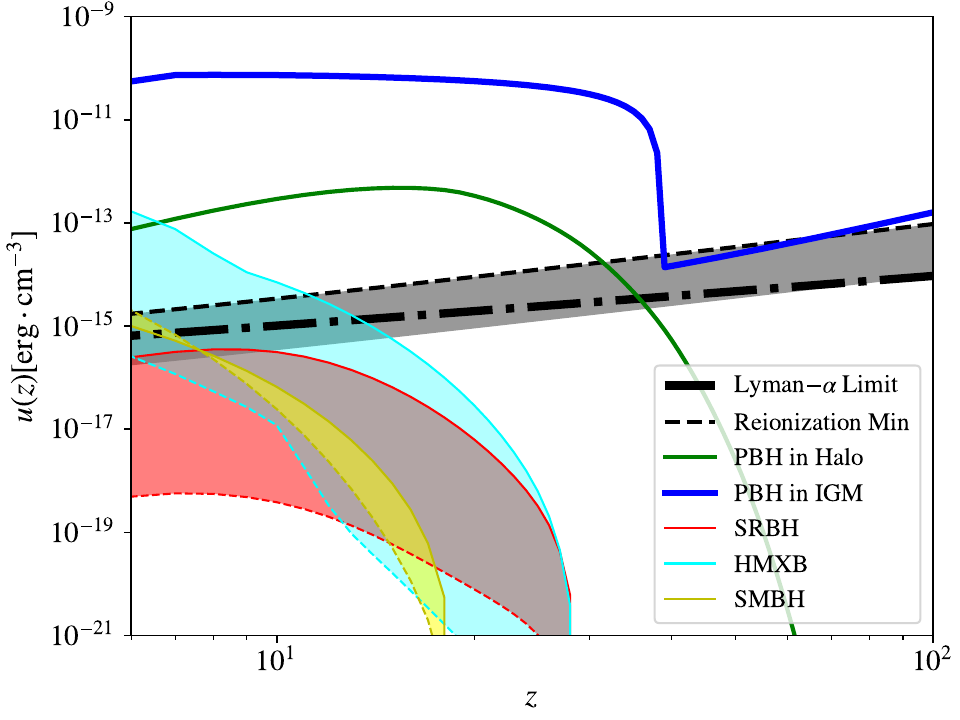}
\end{subfigure}
\hfill
\begin{subfigure}{0.45\textwidth}
    \caption{PBH mass fraction: $f_{\rm PBH} = 10^{-5}$}
    \includegraphics[width=\textwidth]{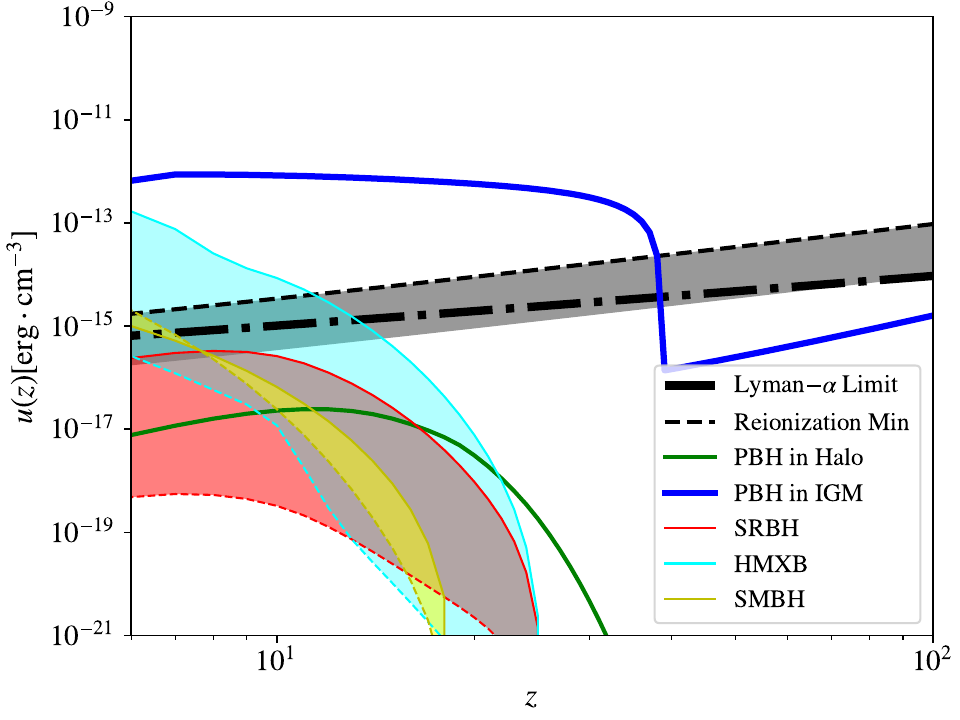}
\end{subfigure}
\caption{Effect of enhanced accretion in PBH-seeded haloes. Similar to Figure~\ref{fig:Ltot}, we show here the integrated (bolometric) energy density from BH accretion feedback of different origins vs. redshift: PBHs in the IGM (blue), in haloes (green), SRBHs (red), HMXBs (cyan), and SMBHs (yellow). For PBHs, we assume a monochromatic mass distribution with $M_{\rm c} = 10 \rm \ M_{\odot}$, and mass fractions of $f_{\rm PBH} = 10^{-3}$ (upper panel) and $10^{-5}$ (lower panel). Here, we consider the same range for the abundance of BHs as shown in Figure~\ref{fig:RhoBH}, when deriving possible radiation outputs, but include the effect of halo enhancement on PBH accretion.}
\label{fig:Ltot_halo}
\end{figure}

Our results, plotted in Figure \ref{fig:Ltot_halo}, are compared with the energy density from PBHs in Figure \ref{fig:Ltot}. We find that the enhancement in accretion could increase the total energy output by an order of $\sim \mathcal{O}(10^4)$ at $z\lesssim 30$, which would imply more stringent constraints when comparing the intensity output from PBHs to the radiation background limits.

\bsp	
\label{lastpage}
\end{document}